\documentclass[amsmath,amssymb,amsfonts,amsthm,10pt,aps,twocolumn,superscriptaddress,pra]{revtex4-2}

\usepackage{graphicx}
\usepackage{latexsym,euscript,xr}
\usepackage{booktabs}
\usepackage{braket}
\usepackage{mathtools}
\usepackage[nodayofweek]{datetime}
\usepackage[colorlinks=true,citecolor=teal,linkcolor=teal]{hyperref}
\usepackage[table]{xcolor}
\usepackage{color,colortbl,tabularx}
\usepackage{float}
\usepackage[english]{babel}
\usepackage[protrusion=true,expansion=true]{microtype}
\usepackage{tikz} 
\usetikzlibrary{patterns}
\usepackage{amsmath,amssymb,amsfonts,amsthm}
\usepackage{bm}
\usepackage{titletoc}
\usepackage[capitalize]{cleveref}

\usepackage{etoolbox}

\makeatletter
\def\@author@join@#1#2#3{%
  \def\@author{{#1}{#2\separate@note{#3}}}%
}
\def\separate@note#1{%
  \frontmatter@footnote{#1}%
  \def\separate@note##1{%
    \comma@space\frontmatter@footnote{##1}%
  }%
}
\patchcmd{\doauthor}
  {\frontmatter@footnote{#2}}
  {#2}
  {}
  {\errmessage{Patching doauthor failed}}
\makeatother

\newcommand{\reals}{\mathbb{R}}
\newcommand{\complex}{\mathbb{C}}

\newcommand{\Tr}{\mathrm{Tr}}
\newcommand{\N}{\mathfrak{N}}

\newtheorem{theorem}{Theorem}
\newtheorem{lemma}{Lemma}

\newtheorem{corollary}{Corollary}

\newtheorem{definition}{Definition}

\definecolor{stateblue}{RGB}{157,195,230}
\definecolor{modeblue}{RGB}{31,78,121}
\definecolor{gateorange}{RGB}{255,192,0}
\tikzset{
  wire/.style={line width=1pt, line cap=round, black!85},
  discard/.style={line width=1.6pt, line cap=round, black!85},
  gate/.style={fill=gateorange, draw=gateorange!70!black, line width=0.6pt, rounded corners=4pt},
  mode/.style={fill=modeblue, draw=white, line width=0.8pt},
  state/.style={fill=stateblue, draw=modeblue!40, line width=0.6pt},
  frame/.style={dashed, rounded corners=8pt, draw=modeblue!70, line width=0.7pt},
  party/.style={fill=stateblue!35, rounded corners=3pt},
}

\begin{document}

\title{Unifying entanglement and nonclassicality via the Narcowich--Wigner spectrum}

\author{Massimo Frigerio}
\thanks{These authors contributed equally.}
\email{massimo.frigerio@lkb.upmc.fr}
\affiliation{
Laboratoire Kastler Brossel,
Sorbonne Universit\'{e}, CNRS, ENS-Universit\'{e} PSL,  Coll\`{e}ge de France, 4 place Jussieu, F-75252 Paris, France
}

\author{Zacharie Van Herstraeten}
\thanks{These authors contributed equally.}
\affiliation{QAT team, DIENS, \'Ecole normale sup\'erieure, PSL University, CNRS, INRIA, 45 rue d'Ulm, Paris 75005, France}

\author{Hania Herbi}
\affiliation{
ENS-Université PSL, 24, rue Lhomond 75005 Paris - France
}

\author{Jack Davis}
\affiliation{QAT team, DIENS, \'Ecole normale sup\'erieure, PSL University, CNRS, INRIA, 45 rue d'Ulm, Paris 75005, France}

\author{Nuno C.\ Dias}
\affiliation{Grupo de Física Matemática, Departamento de Matemática Instituto Superior Técnico, Av. Rovisco Pais, 1049-001 Lisbon, Portugal}
\affiliation{ISCTE - Instituto Universitário de Lisboa, Av. das Forças Armadas, 1649-026, Lisbon, Portugal}

\author{João N.\ Prata}
\affiliation{Grupo de Física Matemática, Departamento de Matemática Instituto Superior Técnico, Av. Rovisco Pais, 1049-001 Lisbon, Portugal}
\affiliation{ISCTE - Instituto Universitário de Lisboa, Av. das Forças Armadas, 1649-026, Lisbon, Portugal}

\author{Mattia Walschaers}
\affiliation{
Laboratoire Kastler Brossel,
Sorbonne Universit\'{e}, CNRS, ENS-Universit\'{e} PSL,  Coll\`{e}ge de France, 4 place Jussieu, F-75252 Paris, France
} 

\author{Ulysse Chabaud}
\affiliation{QAT team, DIENS, \'Ecole normale sup\'erieure, PSL University, CNRS, INRIA, 45 rue d'Ulm, Paris 75005, France}

\begin{abstract}
    Entanglement and nonclassicality are two fundamental features of quantum mechanics which serve as resources in quantum information processing tasks and are tightly connected by two-mode mixing. In this work, we show that the connection between entanglement and nonclassicality extends well beyond the two-mode mixing setting by considering the case of arbitrary multimode Gaussian evolutions. Using the Narcowich--Wigner spectrum as our main technical tool to quantify single-mode phase-space nonclassicality, we derive sufficient conditions for a single-mode state to generate NPT entanglement after a multimode Gaussian unitary evolution, which both tighten and generalize existing criteria. As an application, we obtain simple single-mode protocols for quantitatively testing the entanglement structure of multimode quantum states over arbitrary bipartitions. Our results identify the Narcowich--Wigner spectrum as a powerful tool for investigating fundamental properties of bosonic quantum states.
\end{abstract}

\maketitle

\section{Introduction}

Quantum technologies rely on harnessing truly quantum features of physical systems to achieve quantum advantages with respect to their classical counterparts~\cite{dowling2003quantum}. This raises a fundamental question: what makes a quantum system nonclassical? Answering this question depends on both the physical setting and the operational task under consideration: different notions of nonclassicality isolate different quantum resources, such as coherence, contextuality, or entanglement~\cite{RevModPhys.91.025001}.

For a single canonical degree of freedom, classical statistical states can be identified as arbitrary probability distributions over phase space.  For a single bosonic mode however, quantum states can only be represented using quasiprobability distributions over phase space such as the Wigner function~\cite{Wigner1932} or the Glauber--Sudarshan $P$ function~\cite{sudarshan1963equivalence, glauber1963coherent}. These descriptions provide a natural framework for characterizing quantum nonclassicality as they are each highly constrained by the uncertainty principle and need not satisfy the positivity condition required of phase-space probabilities. 
The Glauber--Sudarshan $P$ representation makes this comparison particularly explicit by expressing a single-mode quantum state as $\hat \rho=\int_{\mathbb{C}}\mathrm{d}^{2}\alpha P_{\hat \rho}(\alpha)\ket\alpha\!\bra\alpha$,
where the coherent states $\{\ket\alpha\}$ play the role of classical phase-space points. A state $\hat \rho$ is called $P$-classical when $P_{\hat\rho}$ is a bona fide probability distribution, and $P$-nonclassical otherwise. In particular, $P$-classical states are statistical mixtures of coherent states.
Beyond this binary distinction, the \textit{nonclassical depth} quantifies how strongly a state departs from the set of $P$-classical states~\cite{lee1991measure}. It is defined as the minimum amount of Gaussian smoothing required to transform the $P$ distribution into a regular nonnegative distribution. This nonclassical depth has a direct physical interpretation: it measures the minimum amount of classical noise required to erase the nonclassicality of the state~\cite{lee1991measure}.

For composite quantum systems, the tensor-product structure introduces another signature of quantumness: entanglement. A multipartite quantum state is separable if it can be written as a convex mixture of product states, and entangled otherwise~\cite{horodecki2009quantum}. And while entanglement does not provide an exhaustive characterization of multipartite quantumness due to the existence of separable states that exhibit other nonclassical correlations, it is nonetheless a central operational resource for quantum-information protocols~\cite{RevModPhys.91.025001}. A well-known criterion for detecting entanglement is the Positive-Partial-Transpose (PPT) test ~\cite{peres1996,horodecki1996}, wherein a bipartite quantum state with a Negative Partial Transpose (NPT) witnesses the presence of so-called \textit{NPT entanglement} with respect to the bipartition at hand.

The notion of entanglement is conceptually distinct from that of phase-space nonclassicality. However, mode mixing provides a direct relation between these two notions of quantumness: when a single-mode state is mixed with the vacuum on a beam-splitter, or more generally with a $P$-classical auxiliary state, passive linear optics can convert its $P$-nonclassicality into bipartite entanglement; conversely, classical inputs remain separable because a positive input $P$ distribution is mapped to a positive joint distribution over product coherent states~\cite{kim2002entanglement,wang2002theorem}. This observation underlies the notion of entanglement potential, which quantifies single-mode nonclassicality through the amount of entanglement generated at the output of a beam-splitter \cite{asboth2005computable,vogel2014unified}. This operational connection between phase-space nonclassicality and entanglement has since been extended to mixing with arbitrary auxiliary states, in which case the nonclassical depth of the input provides a sufficient condition for entanglement at the output \cite{gholipour2016entanglement}. As a result, nonclassicality at the output of a beam-splitter can serve as a witness of entanglement~\cite{gholipour2016entanglement}, a fact which has been generalized and quantitatively characterized in subsequent works~\cite{PhysRevLett.130.160201,zaw2024certifiable,Liu2026quantumentanglement,callus2026revealing,zaw2026sufficient}.

In this work, we revisit the well-known connection between entanglement and phase-space nonclassicality in two distinct ways. 
First, we significantly generalize the scope of this connection to multimode entanglement generated by arbitrary Gaussian unitaries.
Second, rather than focusing on phase-space negativity as the signature of nonclassicality, we employ the \textit{Narcowich--Wigner} (NW) \textit{spectrum}~\cite{Narcowich1986,narcowich_conditions_1988} as our main technical tool. Informally, this spectrum characterizes the admissible values of Planck's constant for which a phase-space function may be interpreted as representing a valid quantum state~\cite{gosson2025sensitivity}.
Our main result is a derivation, based on the NW spectrum, of a set of witnesses for NPT entanglement present across any multimode bipartition of the input state.  Conversely, these witnesses can be used to characterize single-mode states that will become NPT entangled upon interference with a tailored Gaussian unitary and \textit{any} environment. 
These conditions generalize and are strictly tighter than previous conditions based on nonclassical depth~\cite{gholipour2016entanglement}.  We moreover show how to detect such entanglement using measurements performed only on a single output mode.  Finally, we show that these entanglement witnesses provide quantitative lower bounds on entanglement negativity~\cite{vidal2002computable}, and that they additionally become necessary conditions for Gaussian states.

The rest of the paper is structured as follows: in \cref{sec:background} we provide the necessary background as well as an introduction to the NW spectrum; in \cref{sec:ncdandNW}, we connect the nonclassical depth with the NW spectrum; in \cref{sec:NW-entanglement}, we give our main result which relates the NW spectrum to NPT entanglement after a Gaussian unitary; we use this result in \cref{sec:entdet} to obtain entanglement tests which involve measuring a single mode, and in \cref{sec:entgen} to derive sufficient conditions for entanglement generation after a Gaussian unitary; finally, we conclude with a few open questions in \cref{sec:concl}.

\section{Background}
\label{sec:background}

\subsection{Continuous-variable quantum information}
\label{sec:background_CV}

We briefly recall the elements of continuous-variable quantum information that are used throughout the paper, and refer to Refs.~\cite{braunstein2005quantum,weedbrook2012gaussian,BUCCO} for comprehensive accounts.

\subsubsection{Modes, quadratures and displacements}

We consider $N$ bosonic modes with annihilation operators $\hat a_k$, obeying the commutation relations $[\hat a_j,\hat a_k^\dagger]=\delta_{jk}$. The photon-number operator of mode $k$ is $\hat n_k=\hat a_k^\dagger\hat a_k$, and $\hat n=\sum_{k=1}^N\hat n_k$ is the total photon-number operator. Each mode is also associated with a position quadrature $\hat x_k=(\hat a_k+\hat a_k^\dagger)/\sqrt2$ and a momentum quadrature $\hat p_k=i(\hat a_k^\dagger-\hat a_k)/\sqrt2$. In our convention, these quadratures obey $[\hat x_k,\hat p_k]=i$, which effectively corresponds to setting $\hbar=1$.

We group these quadratures in the more convenient vector notation $\hat{\mathbf r}=(\hat x_1,\hat p_1,\dots,\hat x_N,\hat p_N)^T$, ordered mode by mode, whose commutation relations read $[\hat{\mathbf r},\hat{\mathbf r}^T]=i\bm\Omega$, where the left-hand side denotes the matrix of commutators $[\hat r_j,\hat r_k]$ and where
\begin{align}
    \bm\Omega=\bigoplus_{k=1}^N\bm J,
    \qquad
    \bm J=\begin{pmatrix}0&1\\-1&0\end{pmatrix}
\end{align}
is the symplectic form. A point of the classical phase space of the $N$ modes is written $z=(x_1,p_1,\dots,x_N,p_N)^T\in\reals^{2N}$, with the same ordering.

For a single mode, the displacement operator associated with the phase-space point $(x,p)$ is the unitary $\hat D(x,p)=\exp[i(p\hat x-x\hat p)]$, which shifts $\hat x$ by $x$ and $\hat p$ by $p$.
In terms of the mode operators, it is given by $\exp(\alpha\hat a^\dagger-\alpha^*\hat a)$ with $\alpha=(x+ip)/\sqrt2$. 
Its multimode version, associated with a point $z\in\reals^{2N}$, is
\begin{align}
    \hat D(z)
    =\exp\!\big(i\,\hat{\mathbf r}^T\bm\Omega\,z\big)
    =\bigotimes_{k=1}^N\hat D(x_k,p_k),
    \label{eq:displacement}
\end{align}
and acts on the quadratures as $\hat D(z)^\dagger\,\hat{\mathbf r}\,\hat D(z)=\hat{\mathbf r}+z$.

\subsubsection{Quantum states and characteristic functions}

A state over $N$ modes is a density operator $\hat\rho$, that is, a positive semidefinite trace-class operator of unit trace. 
In particular, we will often use the Fock states, defined as eigenstates of the photon-number operator, i.e.\ $\hat{n}\ket{n}=n\ket{n}$ for a single mode.
The Fock state $\ket{0}$ (the vacuum state) can be used to generate the family of coherent states, defined as $\ket{\alpha}=\hat{D}(x,p)\ket{0}$ with $\alpha=(x+ip)/\sqrt{2}$.

The first and second moments of a state are its mean vector and covariance matrix,
\begin{align}
    \bm d=\langle\hat{\mathbf r}\rangle,
    \qquad
    V_{jk}=\tfrac12\big\langle\{\hat r_j-d_j,\hat r_k-d_k\}\big\rangle ,
    \label{eq:covariance}
\end{align}
and the uncertainty principle takes the form of the matrix inequality~\cite{simon1994quantum}
\begin{equation}
    \bm V+\tfrac{i}{2}\,\bm\Omega\succeq0 ,
    \label{eq:uncertainty}
\end{equation}
equivalently $\nu_k\ge\tfrac12$ for the symplectic eigenvalues $\nu_k$ of $\bm V$, the moduli of the eigenvalues of $i\bm\Omega\bm V$~\cite{williamson1936algebraic,weedbrook2012gaussian}; for a single mode, $\nu=\sqrt{\det\bm V}$.
A general state is not characterized by its first two moments, however, and the moments of all orders are conveniently gathered in a single function.

The characteristic function of a trace-class operator, and in particular of a quantum state $\hat{\rho}$, is defined as
\begin{align}
    \chi_{\hat\rho}(z)=\Tr\big[\hat\rho\,\hat D(z)\big],
    \qquad z\in\reals^{2N},
    \label{eq:chi_def}
\end{align}
and determines $\hat\rho$ completely~\cite{cahill1969density}.

A Gaussian state is a state whose characteristic function is Gaussian,
\begin{align}
    \chi_{\hat\rho}(z)=\exp\!\big(-\tfrac12\,z^T\bm\Omega^T\bm V\bm\Omega\,z+i\,\bm d^T\bm\Omega\,z\big),
    \label{eq:gaussian_chi}
\end{align}
and is therefore entirely specified by $\bm d$ and $\bm V$; the vacuum and the coherent states are Gaussian states with $\bm V=\tfrac12\bm I$. 

\subsubsection{Gaussian unitaries}

Gaussian unitaries, generated by Hamiltonians at most quadratic in the quadratures, are characterized by their affine action on the quadratures in the Heisenberg picture,
\begin{align}
    \hat U_G^\dagger\,\hat{\mathbf r}\,\hat U_G=\bm S\,\hat{\mathbf r}+\bm\delta ,
    \label{eq:gaussian_unitary}
\end{align}
where $\bm\delta\in\reals^{2N}$ and $\bm S$ is a real $2N\times2N$ matrix which must preserve the commutation relations, that is, be symplectic: $\bm S\bm\Omega\bm S^T=\bm\Omega$. 
The mean vector and covariance matrix transform accordingly as $\bm{d}\mapsto\bm{S}\bm{d}+\bm{\delta}$ and $\bm V\mapsto\bm S\bm V\bm S^T$, and Gaussian states are mapped to Gaussian states. 
The displacement $\bm\delta$ will play no role in this work: displacements are local unitaries, which affect neither the entanglement of a state nor, as will be clear from its definition, the NW spectrum of its modes. 
We shall therefore set $\bm\delta=0$ from \cref{sec:NW-entanglement} on, and a Gaussian unitary is then fully described by a symplectic matrix.

In terms of the mode operators, Eq.~\eqref{eq:gaussian_unitary} with $\bm\delta=0$ reads
\begin{align}
    \hat b_k\coloneqq\hat U_G^\dagger\,\hat a_k\,\hat U_G=\sum_{l=1}^N\big(A_{kl}\,\hat a_l+B_{kl}\,\hat a_l^\dagger\big),
    \label{eq:bogoliubov}
\end{align}
a so-called Bogoliubov transformation. A Gaussian unitary is passive when $\bm B=0$, in which case it conserves the total photon number and can be implemented with linear optics; otherwise it is active and requires squeezing, the elementary example being the single-mode squeezer $\hat S(r)=\exp[\tfrac r2(\hat a^2-\hat a^{\dagger2})]$, which rescales $\hat x$ by $e^{-r}$ and $\hat p$ by $e^{r}$. Two two-mode Gaussian unitaries, one passive and one active, play a prominent role in this work.

\emph{Beam-splitter.} $\hat U_{\mathrm{BS}}(\theta)=\exp[\theta(\hat a_1^\dagger\hat a_2-\hat a_1\hat a_2^\dagger)]$ mixes two modes passively,
\begin{align}
    \hat b_1&=\cos\theta\,\hat a_1+\sin\theta\,\hat a_2,\\
    \hat b_2&=-\sin\theta\,\hat a_1+\cos\theta\,\hat a_2,
    \label{eq:BS}
\end{align}
and is characterized by its transmittance $T=\cos^2\theta\in[0,1]$.

\emph{Two-mode squeezer.} $\hat U_{\mathrm{TMS}}(r)=\exp[r(\hat a_1^\dagger\hat a_2^\dagger-\hat a_1\hat a_2)]$, $r>0$, is the simplest active two-mode unitary,
\begin{align}
    \hat b_1&=\cosh r\,\hat a_1+\sinh r\,\hat a_2^\dagger,
    \nonumber\\
    \hat b_2&=\cosh r\,\hat a_2+\sinh r\,\hat a_1^\dagger,
    \label{eq:TMS}
\end{align}
and is characterized by its gain $g=\cosh^2 r\ge1$.

\subsection{PPT criterion for entanglement}
\label{sec:background_PPT}

A state is separable across a bipartition $A|B$ of the modes if it is a convex mixture of product states, $\hat\rho=\sum_ip_i\,\hat\rho_A^{(i)}\otimes\hat\rho_B^{(i)}$, and entangled otherwise. Partial transposition on $B$, defined by $(\hat X_A\otimes\hat Y_B)^{\Gamma_B}=\hat X_A\otimes\hat Y_B^{T}$ and extended by linearity, preserves the trace and satisfies $\Tr[\hat X^{\Gamma_B}\hat Y]=\Tr[\hat X\hat Y^{\Gamma_B}]$. Since it maps product states to product states, which are positive operators, a separable state has a positive partial transpose,
\begin{equation}
    \hat\rho\ \text{separable}\quad\Rightarrow\quad\hat\rho^{\Gamma_B}\succeq0 .
    \label{eq:PPT}
\end{equation}
This is the PPT criterion~\cite{peres1996,horodecki1996}: a state whose partial transpose is not positive semidefinite is entangled, and is called NPT-entangled, for negative partial transpose. The converse of Eq.~\eqref{eq:PPT} fails in general, and PPT-entangled states escape the criterion.

In the position representation, transposition leaves $\hat x$ unchanged and reverses $\hat p$. Partial transposition on $B$ thus flips the sign of the momenta of $B$: it maps $\hat{\mathbf r}$ to $\bm\Lambda_B\hat{\mathbf r}$, where 
\begin{equation}
    \bm\Lambda_B=\bigoplus_{k=1}^N
    \begin{cases}
        \bm I_2 & k\in A,\\
        \mathrm{diag}(1,-1) & k\in B.
    \end{cases}
    \label{eq:LambdaB}
\end{equation}
Accordingly, under $\hat\rho\mapsto\hat\rho^{\Gamma_B}$ the mean vector and covariance matrix transform as $\bm d\mapsto\bm\Lambda_B\bm d$ and $\bm V\mapsto\bm\Lambda_B\bm V\bm\Lambda_B$.

For a Gaussian state, $\hat\rho^{\Gamma_B}$ is thus again a Gaussian operator, with covariance matrix $\bm\Lambda_B\bm V\bm\Lambda_B$, and it is a state if and only if this matrix satisfies Eq.~\eqref{eq:uncertainty}. 
Hence, a Gaussian state is PPT across $A|B$ if and only if
\begin{equation}
    \bm V+\tfrac{i}{2}\,\bm\Lambda_B\bm\Omega\bm\Lambda_B\succeq0 ,
    \qquad
    \bm\Lambda_B\bm\Omega\bm\Lambda_B=\bm\Omega_A\oplus(-\bm\Omega_B),
    \label{eq:simon}
\end{equation}
where $\bm\Omega_A$ and $\bm\Omega_B$ are the symplectic forms of the two parties: reflecting the momenta of $B$ reverses its commutation relations, and the partial transpose of a Gaussian state is a state exactly when the uncertainty principle still holds with the commutation relations of $B$ reversed. 
This is Simon's criterion~\cite{Simon_PPT_2000_PRL} and its multimode extension~\cite{werner2001bound}. 
When one of the two parties is a single mode, Eq.~\eqref{eq:simon} is also sufficient for separability~\cite{Simon_PPT_2000_PRL,werner2001bound}, whereas for $|A|,|B|\ge2$ there exist PPT-entangled Gaussian states~\cite{werner2001bound}.

\subsection{Phase-space representations and NW spectrum}
\label{subsec:phase-space-and-NW}

Continuous-variable states admit a convenient description in terms of functions on phase space, the most common of which is the Wigner function~\cite{Wigner1932}. 
It is the symplectic Fourier transform of the characteristic function,
\begin{align}
    W_{\hat\rho}(z)
    =\frac{1}{(2\pi)^{2N}}\int_{\reals^{2N}}\chi_{\hat\rho}(\xi)\,e^{-i\,z^T\bm\Omega\,\xi}\,\mathrm d^{2N}\xi ,
    \label{eq:wigner}
\end{align}
and conversely $\chi_{\hat\rho}(\xi)=\int W_{\hat\rho}(z)\,e^{i\,z^T\bm\Omega\,\xi}\,\mathrm d^{2N}z$. 

The Wigner function is real and normalized, its marginals are the probability densities of the quadratures, and expectation values are phase-space averages, $\Tr[\hat X\hat Y]=(2\pi)^N\!\int W_{\hat X}W_{\hat Y}$, where $W_{\hat X}$ is defined by Eq.~\eqref{eq:wigner} for any trace-class operator $\hat X$. 
In all these respects it behaves like the probability distribution of a classical state, and it is indeed a probability distribution for Gaussian states, whose Wigner function is a Gaussian. 
It may however take negative values, hence the name quasiprobability distribution, and this negativity is the most familiar signature of nonclassicality in phase space.

\subsubsection{s-parametrised quasiprobability distributions}
The Glauber--Sudarshan $P$ function, the Wigner function, and the Husimi $Q$ function are particular members of the one-parameter family of quasiprobability distributions introduced by Cahill and Glauber \cite{cahill1969density}. For a single-mode quantum state $\hat{\rho}$, this family is denoted by
\begin{equation}
    \left\{W_{\hat{\rho}}^{(s)}(\alpha)\right\}_{s\in[-1,1]},
\end{equation}
where the ordering parameter $s$ specifies the operator ordering associated with the corresponding phase-space representation. The specific cases
\begin{align}
W_{\hat{\rho}}^{(1)}(\alpha)=P_{\hat{\rho}}(\alpha)  \\
W_{\hat{\rho}}^{(0)}(\alpha)=W_{\hat{\rho}}(\alpha) \\
W_{\hat{\rho}}^{(-1)}(\alpha)=Q_{\hat{\rho}}(\alpha),
\end{align}
correspond, respectively, to the Glauber--Sudarshan, Wigner, and Husimi representations.
For any pair of ordering parameters $s_2<s_1$, the corresponding quasiprobability distributions are related by the Gaussian convolution
\begin{equation}\label{eq:froms1tos2}
W_{\hat{\rho}}^{(s_2)}(\alpha)
\!=\!
\frac{2}{\pi(s_1-s_2)}
\int_{\mathbb C}\!
W_{\hat{\rho}}^{(s_1)}(\beta)
\exp\!\left[
-\frac{2|\alpha-\beta|^2}{s_1-s_2}
\right]
\!\mathrm d^2\beta,
\end{equation}
which maps the representation with parameter $s_1$ onto a more regular representation with parameter $s_2$. Thus, decreasing $s$ corresponds to increasing the amount of Gaussian smoothing applied to the phase-space distribution. The inverse transformation is a Gaussian deconvolution and, in general, does not define a regular function, consistent with the fact that the Glauber--Sudarshan $P$ representation can be singular.
\newline

The parameter $s$ thus induces a natural hierarchy of phase-space representations. The Glauber--Sudarshan distribution ($s=1$) may fail to be a regular function and can exhibit highly singular behavior. At the opposite end of the hierarchy, the Husimi function ($s=-1$) is obtained after maximal Gaussian smoothing and is a normalized, continuous, and everywhere nonnegative function for every density operator. The Wigner function ($s=0$) occupies an intermediate position: it is well defined as a phase-space distribution for trace-class states, but may take negative values, providing a signature of nonclassicality. This continuous interpolation between the $P$, Wigner, and $Q$ representations provides a natural framework for quantifying nonclassicality through Lee's notion of nonclassical depth \cite{lee1991measure}.

\subsubsection{Nonclassical depth}

The continuous interpolation provided by the Cahill--Glauber family naturally leads to a quantitative measure of nonclassicality. Since decreasing the ordering parameter $s$ corresponds to increasing Gaussian smoothing of the phase-space representation, one may ask how much smoothing is required before the quasiprobability distribution becomes an everywhere nonnegative function.

A quantum state is said to be \textit{classical} if its Glauber--Sudarshan $P$ representation is a regular, everywhere nonnegative probability density.  
This notion of classicality intuitively stems from electromagnetic explainability: the optical statistics of $P$-positive states admit a description as statistical averages over solutions of Maxwell’s equations, whereas those of $P$-negative states do not.
Within the Cahill--Glauber family, one may determine the smallest amount of Gaussian smoothing required for the corresponding $s$-parametrized quasiprobability distribution to become everywhere nonnegative. This motivates the definition of the critical ordering parameter of a state $\hat\rho$:
\begin{equation}
s_{\hat\rho}
=
\sup\left\{
s\in[-1,1]
\,\middle|\,
W_{\hat{\rho}}^{(s)}(\alpha)\geq 0
\ \text{for all }\alpha\in\mathbb{C}
\right\},
\end{equation}
which is well defined because the Husimi function
\begin{equation}
Q_{\hat{\rho}}(\alpha)=W_{\hat{\rho}}^{(-1)}(\alpha),
\end{equation}
is everywhere nonnegative. The parameter $s_{\hat\rho}$ identifies the least smoothed member of the Cahill--Glauber family that is everywhere nonnegative.

Following Lee \cite{lee1991measure}, the nonclassical depth of a quantum state $\hat{\rho}$ is defined as
\begin{equation}\label{nonclassical_depth_def}
\tau_{\hat{\rho}}
=
\frac{1-s_{\hat\rho}}{2},
\end{equation}
with $\tau_{\hat{\rho}}\in[0,1]$. The nonclassical depth quantifies the minimum amount of Gaussian smoothing required to render the Glauber--Sudarshan function of $\hat \rho$ everywhere nonnegative. In particular, $\tau_{\hat{\rho}}=0$ when the Glauber--Sudarshan $P$ representation is already a regular, nonnegative probability density, as for coherent states. Conversely, $\tau_{\hat{\rho}}=1$ when positivity is attained only at $s=-1$, i.e., at the level of the Husimi $Q$ function. Such states are maximally nonclassical according to Lee's criterion.

\subsubsection{The Narcowich--Wigner spectrum}
\label{sec:background_NW}

Finally, we provide background on the Narcowich--Wigner spectrum.
Let $F$ be some complex-valued function defined on $\mathbb{R}^{2N}$ and let $\tilde{F}$ denote its symplectic Fourier transform
\begin{equation}
\tilde{F}(\xi)=\int_{\mathbb{R}^{2N}}F(z)e^{iz^T\bm\Omega \xi}dz~. 
\end{equation}
Given $\eta \in \mathbb{R}$, we say that $\tilde{F}$  is of the $\eta$-\textit{positive type}
if 

\begin{enumerate}
    \item $\tilde{F}(\xi)$ is continuous, and
    \item for any $m \in \mathbb{N}$ and any set of $m$ vectors $\xi_1, \cdots, \xi_m \in \mathbb{R}^{2N}$ the matrix $\bm M(\eta)$ with entries $\tilde{F}(\xi_j-\xi_k)e^{\frac{i\eta}{2}\xi_j^T\bm\Omega \xi_k}$, $j,k=1, \cdots,m$ is positive semi-definite.
\end{enumerate}

The \textit{Narcowich-Wigner spectrum} (NW) of $F$ is the set \cite{cordero2019,broeckerwerner1995,narcowich_conditions_1988,Narcowich1986,diasprata2018}:
\begin{equation}
   \N(F)=\left\{\eta \in \mathbb{R}:~\tilde{F} \text{ is of the $\eta$-positive type}\right\}. 
\label{eqNWspectrum1}
\end{equation}
According to Bochner's theorem the functions of $0$-positive type are exactly those whose Fourier transform is non-negative. And a theorem by Kastler \cite{Kastler1965} and by Loupias and Miracle-Solé~\cite{Loupias1966,Loupias1967} states that a function $F$ is the Wigner function of some state $\hat{\rho}$ if and only if it satisfies the so-called \textit{KLM conditions}:

\begin{enumerate}
    \item $\tilde{F}(0)=1$, and
\item $\tilde{F}$ is of the $1$-positive type.
\end{enumerate}

This can be summarized by saying that, if $F$ is everywhere non-negative then $0\in \N(F)$, and if $F$ is the Wigner function of a valid quantum state, then $1 \in \N(F)$.
For a generic $\eta \in \N(F)$, $\eta$-positivity can thus be understood in the following sense: there exists a positive trace-class operator $\hat{\rho}$ such that $F(z)=|\eta|^{-N} W_{\hat\rho}(z/\sqrt{|\eta|})$, the Wigner function where Planck's constant $\hbar=1$ is replaced by $|\eta|$.

In Appendix \ref{app:NWproperties}
we review the properties of NW spectra, and we mention here the main ones used throughout this work.
For every real-valued phase-space function $F$, the NW spectrum is symmetric, 
\begin{equation} 
\eta\in\N(F) \quad\Longleftrightarrow\quad -\eta\in\N(F).
\end{equation} 
Under the usual regularity assumptions, namely $F\in L^2(\mathbb R^{2N})$ and $\widetilde F$ continuous, $\N(F)$ is a compact subset of $\mathbb R$. 
Moreover, if $F*G$ is well defined, then 
\begin{equation} 
\label{eq:NWconvolution}
\N(F)+\N(G)\subseteq \N(F*G), 
\end{equation} 
where the sum on the left-hand side is the Minkowski sum. In particular, if $\eta\in\N(F)$ and $\eta'\in\N(G)$, then $\eta+\eta'\in\N(F*G)$. 

Finally, if $F$ is a normalized Gaussian function with covariance matrix $\bm V$, then 
\begin{equation} 
\label{eq:NWGaussian}
\eta\in\N(F) \quad\Longleftrightarrow\quad \bm V+\frac{i\eta}{2}\bm\Omega\succeq0. 
\end{equation} 
Consequently, if $\nu_{\min}(\bm V)$ denotes the smallest symplectic eigenvalue of $\bm V$, then 
\begin{equation} \N(F) = [-2\nu_{\min}(\bm V),2\nu_{\min}(\bm V)]. 
\end{equation} 
See, for instance, \cite{narcowich_conditions_1988,diasprata2009}.
In particular, $F$ represents a valid quantum Gaussian state when $\nu_{\min}(\bm V)\geq \frac12$, or equivalently $1\in\N(F)$.

Hereafter, for convenience, for an operator $\hat A$, $\N(\hat A)$ refers to the NW spectrum of $W_{\hat A}$.

\section{Nonclassicality and Narcowich--Wigner spectrum}
\label{sec:ncdandNW}

Having introduced the necessary background, we now relate the NW spectrum and the nonclassical depth. As we shall see, these two notions are directly related through the convolution structure of the Cahill--Glauber family.

Let $W_{\hat{\rho}}$ denote the Wigner function of a single-mode quantum state $\hat{\rho}$. For the choice $s_1=0$ and $s_2=-s$, with $s\in[0,1]$ in \cref{eq:froms1tos2}, the $s$-ordered quasiprobability distribution is obtained from the Wigner function by Gaussian convolution,
\begin{equation}
    W^{(-s)}_{\hat{\rho}}(\alpha) = \frac{2}{\pi s} \iint W_{\hat{\rho}}(\beta) e^{-2 \frac{|\alpha -\beta|^2}{s}}\mathrm d^2\beta,
\end{equation}

Since the Gaussian kernel has a NW spectrum containing the interval $[-s,s] $ (see \cref{eq:NWGaussian}), the convolution property \cref{eq:NWconvolution} implies
\begin{equation}
    \N\!\left(W_{\hat{\rho}}^{(-s)}\right) \supseteq \N(W_{\hat{\rho}}) + [-s,s] = \!\!\!\bigcup_{\eta \in \N(W_{\hat \rho})}\!\!\![\eta - s, \eta + s].
\end{equation}
That is, the convolution by a Gaussian with width $s$ implies a thickening of each existing point in the NW spectrum of $W_{\hat \rho}$ into an interval of width $2s$.  If any one of these guaranteed intervals in the thickened spectrum contains zero, then $W_{\hat \rho}^{(-s)}$ is necessarily non-negative as a probability distribution.  Equivalently by contraposition, if $W_{\hat \rho}^{(-s)}$ remains negative after the convolution, then none of the intervals contain zero in the thickened spectrum.  As a result,
\begin{equation}
    W^{(-s)}_{\hat \rho}(\alpha) \ngeq 0 \quad \Rightarrow \quad \N(W_{\hat{\rho}})
    \cap[-s,s]
    =\varnothing.
\end{equation}
But by definition of $s_{\hat\rho}$ being the largest ordering parameter for which $W_{\hat \rho}^{(s)}$ is everywhere non-negative, the preceding implication is equivalent to a bound on the nonclassical depth \eqref{nonclassical_depth_def}:
\begin{equation}\label{eq:ncdandNW}
    \tau_{\hat{\rho}}
    >
    \frac{1+s}{2}
    \quad\Longrightarrow\quad
    \N(W_{\hat{\rho}})
    \cap[-s,s]
    =
    \varnothing,
\end{equation}
for all $0 \leq s < -s_{\hat\rho}$, provided that $s_{\hat\rho}<0$. In other words, a sufficiently large nonclassical depth necessarily implies a gap around the origin in the NW spectrum of the original function $W_{\hat \rho}$.

\begin{figure}[t]
    \centering
    \includegraphics[width=\linewidth]{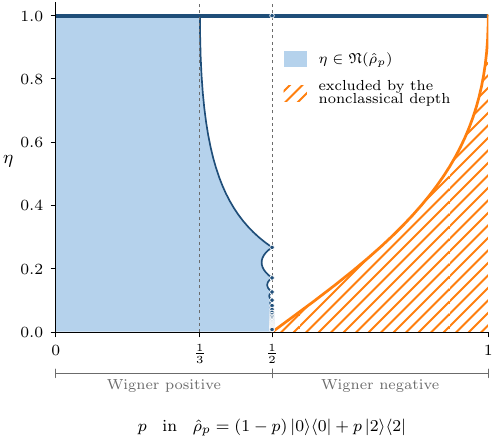}
    \caption{NW spectrum versus nonclassical depth for $\hat\rho_p=(1-p)\ket0\!\bra0+p\ket2\!\bra2$ (blue region adapted from Fig.~3 of Ref.~\cite{broeckerwerner1995}, where the spectrum of this family was computed). Blue: positive half of $\N(\hat\rho_p)$, which is symmetric and always contains $\pm1$ (thick line). It equals $[-1,1]$ for $p\le\tfrac13$. For $\tfrac13<p<\tfrac12$ the state is Wigner positive but its spectrum has gaps, then fragments further, until at $p=\tfrac12$ it collapses to the countable set (dots). For $p>\tfrac12$ it reduces to $\{\pm1\}$. Hatched: interval $|\eta|<2\tau(p)-1$ excluded by the nonclassical depth; it is empty for $p\le\tfrac12$, although $\tau>0$ for all $p>0$. White: values excluded by the NW spectrum alone; by \cref{thm:single_mode}, each of them witnesses NPT entanglement that the nonclassical depth cannot detect.}
    \label{fig:nw_vs_ncd}
\end{figure}

In particular, if $\tau_{\hat{\rho}}>\frac{1+s}{2}, $ then no point of the NW spectrum lies in the interval $[-s,s]$. For states with
\begin{equation}
    \tau_{\hat{\rho}}\leq\frac{1}{2},
\end{equation}
equivalently $s_{\hat\rho}\geq0$, the spectrum therefore has no nonzero interval around the origin that can be inferred solely from positivity of the corresponding $s$-ordered quasiprobability distribution. The limiting case $\tau_{\hat{\rho}}=1$ corresponds to $s_{\hat\rho}=-1$, for which positivity is reached only at the Husimi $Q$-function.

Importantly, the implication \eqref{eq:ncdandNW} is not an equivalence, so the NW spectrum can have values of $\eta$ missing beyond those implied by the nonclassical depth. 
\Cref{fig:nw_vs_ncd} illustrates this with the example of a mixture of vacuum and two photons. 
For a range of mixing parameters the state is Wigner-positive, so that $\tau\le\frac12$ and Eq.~\eqref{eq:ncdandNW} does not exclude values from the NW spectrum, yet its NW spectrum has a gap. 
As we show in \cref{sec:NW-entanglement}, every value missing from the NW spectrum of a single-mode state is a signature of entanglement: such a Wigner-positive state, sent through a suitable beam-splitter, is entangled with whatever enters the other port, a fact invisible to the nonclassical depth and to any criterion based on Wigner negativity \cite{gholipour2016entanglement}. 

The NW spectrum and the nonclassical depth thus provide complementary descriptions of phase-space nonclassicality. The nonclassical depth quantifies the amount of Gaussian smoothing required to obtain an everywhere nonnegative quasiprobability distribution, whereas the NW spectrum characterizes the rescaling parameters for which the underlying phase-space function satisfies a generalized KLM positivity condition. The relation derived above shows that a sufficiently large nonclassical depth enforces a spectral gap around $\eta=0$. Conversely, the presence of spectral points arbitrarily close to the origin guarantees positivity after a corresponding amount of Gaussian smoothing. The structure of the NW spectrum therefore encodes quantitative information about the robustness of nonclassical phase-space structure under Gaussian smoothing.

\subsection*{Examples}

For illustrative purposes, we now provide explicit examples of families of states for which we compute the nonclassical depth and the NW spectrum.

\subsubsection{Single-photon-subtracted squeezed thermal states with probabilistic success}\label{1PSSTS}

We first consider a state $\hat\rho_q$ obtained via a probabilistic photon-subtraction applied to a squeezed thermal state. Hereafter, the parameter $q\in[0,1]$ denotes the probability of failure, such that the resulting state is a statistical mixture of the unconditioned squeezed thermal state and the photon-subtracted component. We write
\begin{equation}
\hat{\rho}_q
=
q\,\hat{\rho}_{\mathrm{STS}}
+
(1-q)\,\hat{\rho}_{\mathrm{PSSTS}},
\label{eq:probabilistic_state}
\end{equation}
where
\begin{align}
\hat{\rho}_{\mathrm{STS}}
&=
\hat{S}(r)\hat{\rho}_{\mathrm{th}}\hat{S}^{\dagger}(r),\\
\hat{\rho}_{\mathrm{PSSTS}}
&=
\frac{
\hat{a}\hat{\rho}_{\mathrm{STS}}\hat{a}^{\dagger}
}{\mathcal{N}},
\qquad
\mathcal{N}
=
\frac{\nu\cosh(2r)-1}{2}.
\end{align}
Here, $\nu=2\bar{n}+1$ parametrizes the thermal occupation, with $\bar{n}$ the mean thermal photon number, and $r$ denotes the squeezing parameter.

The Wigner function of the resulting state is the corresponding convex combination,
\begin{equation}
W_q(x,p)
=
q\,W_{\mathrm{STS}}(x,p)
+
(1-q)\,W_{\mathrm{PSSTS}}(x,p).
\label{eq:probabilistic_wigner_mixture}
\end{equation}
Since the state before photon subtraction is Gaussian, Wigner negativity can only arise from the photon-subtracted component. In the parameter regime $\nu<\cosh(2r)$, the mixture retains Wigner negativity provided that the failure probability is sufficiently small. More precisely, negativity occurs for
\begin{equation}
0\leq q<q_{\max},
\qquad
q_{\max}
=
\frac{\cosh(2r)-\nu}
{\cosh(2r)(1+\nu^2)-2\nu}.
\label{eq:qmax_negativity}
\end{equation}
For $\nu\geq\cosh(2r)$, the Wigner function is non-negative for every $q\in[0,1]$.

\paragraph{Nonclassical depth.}

To quantify the robustness of the nonclassical phase-space structure against Gaussian smoothing, we consider the $s$-ordered quasiprobability distribution
$W_q^{(-s)}$, obtained by convolution with a Gaussian of width $s$. 

Its sign is that of its value at the phase-space origin, so the minimal smoothing parameter that makes it nonnegative is determined by
\begin{equation}
W_q^{(-s_{\max})}(0,0)=0.
\label{eq:probabilistic_depth_condition}
\end{equation}
We denote the resulting critical smoothing parameter by $s_{\max}=s_{\max}(\nu,r,q)$, with the explicit expression given in Appendix~\ref{app:probabilistic_PSSTS}; for $q<q_{\max}$, the nonclassical depth is $\tau_{\hat\rho_q}=(1+s_{\max})/2$, so that Eq.~\eqref{eq:ncdandNW} excludes the interval $|\eta|<s_{\max}$ from the NW spectrum.
The dependence on $q$ quantifies the degradation of nonclassicality induced by imperfect photon subtraction.

\paragraph{Narcowich--Wigner spectrum.}

We now characterize the NW spectrum of the probabilistic state, which determines the values of the symplectic parameter for which its characteristic function is of positive type. After undoing the squeezing and rescaling the phase-space variables by $1/\sqrt{|\eta|}$, the characteristic function takes the Gaussian-polynomial form
\begin{equation}
\chi_q(\tilde{x},\tilde{p})
\propto
e^{-\frac{\nu}{4|\eta|}(\tilde{x}^2+\tilde{p}^2)}
\left[
1-A_q\tilde{x}^2-B_q\tilde{p}^2
\right],
\label{eq:probabilistic_characteristic}
\end{equation}
where $A_q$ and $B_q$ are non-negative coefficients determined by the state parameters and the probabilistic preparation protocol. Their explicit expressions are provided in Appendix~\ref{app:probabilistic_PSSTS}.

\begin{figure}[t]
    \centering
    \includegraphics[width=1\linewidth]{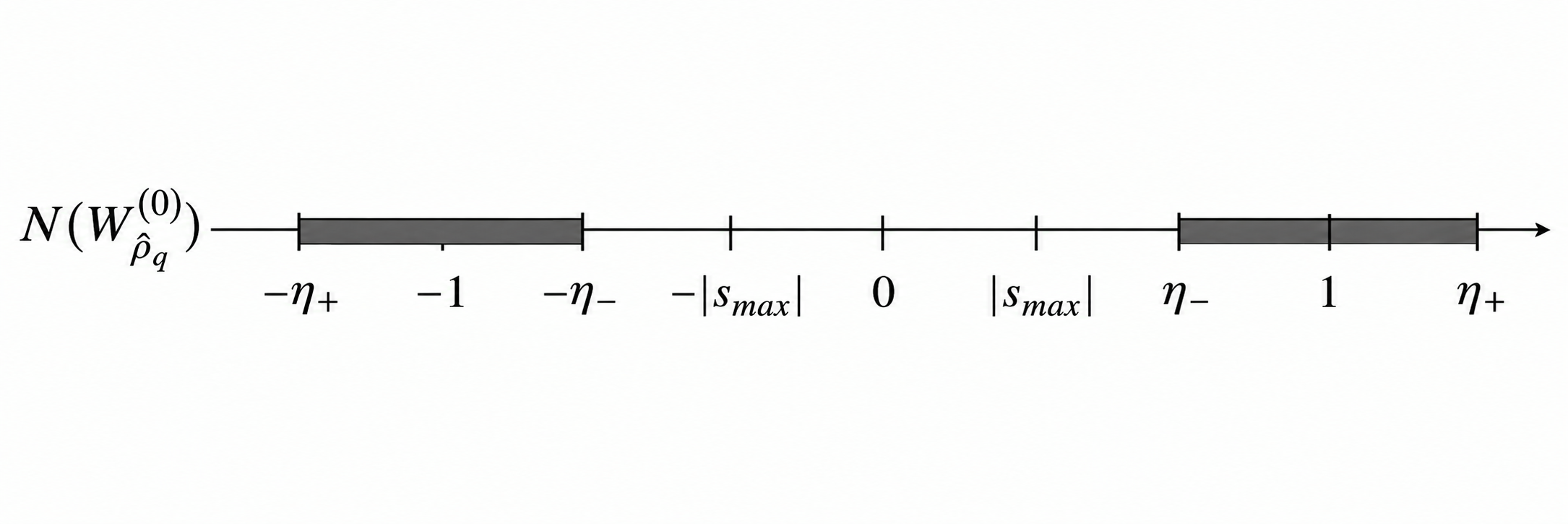}
    \caption{Schematic depiction of the NW spectrum for the single-photon-subtracted squeezed thermal state $\hat\rho_q$.}
    \label{fig:STS}
\end{figure}

The $\eta$-positivity condition can be reduced to the positivity of an effective quadratic operator. Upon a suitable canonical rescaling of the quadratures, this operator takes the form of an isotropic harmonic oscillator plus a constant. Consequently, its positivity is equivalent to the non-negativity of its lowest eigenvalue. This yields a compact characterization of the NW spectrum (see \cref{fig:STS}):
\begin{equation}
|\eta|\in
\left[
\eta_-(\nu,r,q),\,
\eta_+(\nu,r,q)
\right]
\label{eq:probabilistic_NW_interval}
\end{equation}
for $q<q_{\max}$; for $q\geq q_{\max}$ one has $\eta_-\leq0$ and the spectrum is the full interval $[-\eta_+,\eta_+]$.

The explicit expressions for the endpoints are given in Appendix~\ref{app:probabilistic_PSSTS}. Interestingly, the mismatch between $\eta_{-}$ and $s_{\max}$ disappears for $q=0$, i.e.\ for deterministic photon subtraction, meaning that for those states the inner gap of the NW spectrum is fully determined by their nonclassical depth. 

\subsubsection{Two-photon-subtracted single-mode squeezed vacuum state with probabilistic success}
\begin{figure}
    \centering
    \includegraphics[width=\linewidth]{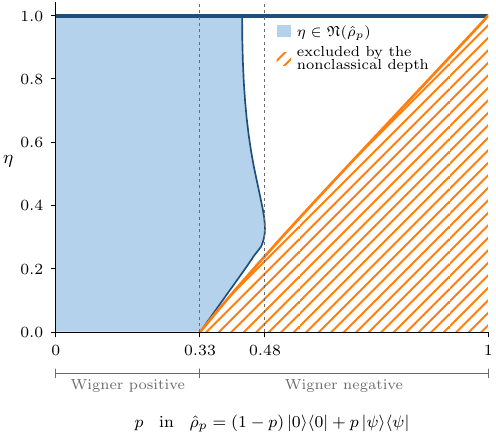}
    
    \caption{
    NW spectrum $\N(\hat{\rho}_p)$ as a function of the photon-subtraction success probability $p$, for the conditional state
    $\hat{\rho}_p=(1-p)|0\rangle\langle 0|+p|\psi\rangle\langle\psi|$,
    with $|\psi\rangle=\alpha|0\rangle+\sqrt{1-\alpha^2}|2\rangle$, $\alpha=(1+2\tanh^2 r)^{-1/2}$, and squeezing parameter $r=2.5$.
    The blue region represents the values of the parameter $\eta$ belonging to the NW spectrum, while the white region indicates excluded values.
    The spectrum exhibits an additional gap on the side of $1$, beyond the central exclusion interval associated with the nonclassical depth (hatched). This illustrates that the NW spectrum can encode nonclassicality-related structure that is not captured by the nonclassical depth alone.
    }
    \label{fig:nw_spectrum}
\end{figure}

Next, we consider a probabilistic $2$-photon subtraction on a single-mode squeezed vacuum state. Upon removing the squeezing after the $2$-photon subtraction, which does not affect the NW spectrum, such a state can be written as:
\begin{equation}
\label{eq:2PSSV}
\begin{aligned}
& \hat{\rho}_{\mathrm{2PSSV}}(p)=(1-p)|0\rangle\langle0|+p|\psi\rangle\langle\psi|,\quad \\
&|\psi\rangle=\alpha|0\rangle+\sqrt{1-\alpha^2}\,|2\rangle,
\end{aligned}
\end{equation}
where $p$ is the probability of success of the two-photon subtraction~\footnote{We assume that either both photons are successfully subtracted, or no photon is subtracted at all.} and \(\alpha = (1 + 2 \tanh^2 r)^{-1/2}\) depends on the squeezing parameter $r$. \cref{fig:nw_spectrum} illustrates its NW spectrum as a function of the photon-subtraction success probability \(p\) for squeezing parameter \(r=2.5\) (see Appendix~\ref{app:probabilistic_2PSSTS} for details on the numerics). This state provides a physically motivated example of how the NW spectrum and the nonclassical depth encode complementary information about nonclassicality, rather than an equivalent one. As shown in \cref{fig:nw_spectrum}, there exist values of $p$ for which the NW spectrum does not reduce to the complement of the nonclassicality interval \([-|s_{\max}|,|s_{\max}|]\) in the interval $[-1,1]$. Although Wigner negativity excludes a symmetric interval around the origin, additional spectral gaps can occur outside this interval: in particular, the plot exhibits a gap close to \(\eta=1\) (and to $-1$ by symmetry of the NW spectrum), which is not captured by nonclassicality alone.

\section{Narcowich--Wigner spectrum and entanglement}
\label{sec:NW-entanglement}

\begin{figure*}
    \centering
    \begin{tikzpicture}[font=\normalsize, txt/.style={black!85}]
  \draw[frame] (-0.9,-3.2) rectangle (6.0,3.15);
  \node[anchor=north west, txt, font=\small] at (-0.75,3.0)
    {\textbf{(a)}\; Generation (\cref{sec:entgen})};
  \fill[fill=black!12, draw=black!30, line width=0.6pt] (0.3,-0.4) ellipse (0.55 and 1.17);
  \foreach \y in {1.2,0.4,-0.4,-1.2} { \draw[wire] (0.3,\y) -- (1.3,\y); }
  \fill[mode] (0.3,1.2) circle (0.29);
  \node[left=5pt, txt] at (0.3,1.2) {$\hat\sigma$};
  \foreach \y in {0.4,-0.4,-1.2} {
    \fill[fill=black!30, draw=white, line width=0.8pt] (0.3,\y) circle (0.29);
  }
  \fill[gate] (1.3,-1.8) rectangle (2.9,1.8);
  \node[txt] at (2.1,0) {$\hat U_G^\dagger$};
  \foreach \y in {1.2,0.4,-0.4,-1.2} { \draw[wire] (2.9,\y) -- (6.0,\y); }
  \fill[state] (4.7,0) ellipse (0.75 and 2.0);
  \foreach \y in {1.2,0.4,-0.4,-1.2} {
    \draw[wire] (3.95,\y) -- (5.45,\y);
    \fill[mode] (4.7,\y) circle (0.29);
  }
  \node[txt] at (4.7,-2.3) {$\hat\rho$};
  \node[txt, font=\small] at (2.55,-2.8)
    {$\hbar_{A|B}(\hat b)\notin\N(\hat\sigma)\ \Rightarrow\ \hat\rho$ NPT-entangled};
  \fill[party] (6.2,0.1) rectangle (7.6,1.5);
  \fill[party] (6.2,-1.5) rectangle (7.6,-0.1);
  \foreach \y in {1.2,0.4,-0.4,-1.2} { \draw[wire, dashed] (6.0,\y) -- (7.8,\y); }
  \node[txt] at (6.9,0.8) {$A$};
  \node[txt] at (6.9,-0.8) {$B$};
  \draw[frame] (7.8,-3.2) rectangle (14.7,3.15);
  \node[anchor=north west, txt, font=\small] at (7.95,3.0)
    {\textbf{(b)}\; Detection (\cref{sec:entdet})};
  \foreach \y in {1.2,0.4,-0.4,-1.2} { \draw[wire] (7.8,\y) -- (10.9,\y); }
  \fill[state] (9.1,0) ellipse (0.75 and 2.0);
  \foreach \y in {1.2,0.4,-0.4,-1.2} {
    \draw[wire] (8.35,\y) -- (9.85,\y);
    \fill[mode] (9.1,\y) circle (0.29);
  }
  \node[txt] at (9.1,-2.3) {$\hat\rho$};
  \fill[gate] (10.9,-1.8) rectangle (12.5,1.8);
  \node[txt] at (11.7,0) {$\hat U_G$};
  \draw[wire] (12.5,1.2) -- (13.5,1.2);
  \fill[mode] (13.5,1.2) circle (0.29);
  \node[right=5pt, txt] at (13.5,1.2) {$\hat\sigma$};
  \foreach \y in {0.4,-0.4,-1.2} {
    \draw[wire] (12.5,\y) -- (13.4,\y);
    \draw[discard] (13.4,\y-0.17) -- (13.4,\y+0.17);
  }
  \node[txt, font=\small] at (11.25,-2.8)
    {$\hat\rho$ PPT $\Rightarrow\ \hbar_{A|B}(\hat b)\in\N(\hat\sigma)$};
\end{tikzpicture}
    \caption{Two consequences of \cref{thm:single_mode}. (a)~Entanglement generation: a single-mode state $\hat\sigma$, together with an arbitrary state of the other modes (grey), is sent through a Gaussian unitary $\hat U_G^\dagger$. If the NW spectrum of $\hat\sigma$ does not contain $\hbar_{A|B}(\hat b)$, the resulting $N$-mode state $\hat\rho$ is NPT-entangled across the bipartition $A|B$ of its modes. (b)~Entanglement detection: $\hat\rho$ is sent through $\hat U_G$, and a single output mode $\hat b$, with reduced state $\hat\sigma$, is retained. If $\hat\rho$ is PPT across $A|B$, the NW spectrum of $\hat\sigma$ contains $\hbar_{A|B}(\hat b)$; conversely, the absence of $\hbar_{A|B}(\hat b)$ from $\N(\hat\sigma)$ certifies NPT entanglement of $\hat\rho$.}
    \label{fig:detection_generation}
\end{figure*}
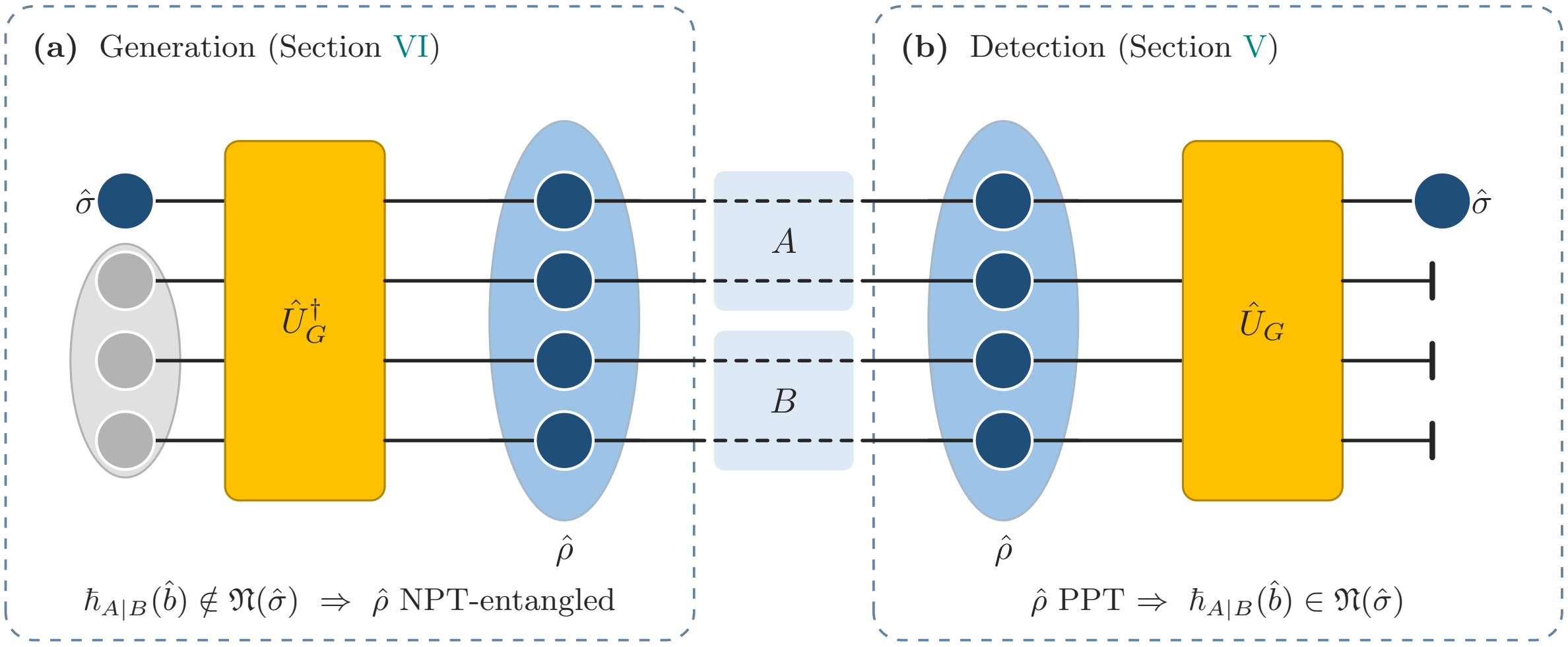

Having clarified the connection between nonclassical depth and NW spectrum, we relate in this section the entanglement of a multimode state to the NW spectrum of a single mode. 

We consider the following setting: an $N$-mode state $\hat\rho$, with a given bipartition $A|B$ of its modes, is sent through a Gaussian unitary $\hat U_G$, and a single output mode $\hat b$ is retained, with reduced single-mode state $\hat\sigma$. 
Our main result (see \cref{thm:single_mode} below) states that if $\hat\rho$ is PPT across $A|B$, then $\N(\hat\sigma)$ contains a value fixed by the unitary and the bipartition alone.

We first introduce a few notations. Let $\hat U_G$ be a Gaussian unitary on $N$ modes and $\hat b$ one of its output modes. 
In the Heisenberg picture, $\hat b$ is a linear combination of the input annihilation and creation operators,
\begin{align}
    \hat b=\sum_{k=1}^N\big(u_k\hat a_k+v_k\hat a_k^\dagger\big),
    \qquad
    [\hat b,\hat b^\dagger]=\sum_{k=1}^N w_k=1 ,
    \label{eq:b}
\end{align}
where
\begin{align}
    w_k\coloneqq|u_k|^2-|v_k|^2=[\hat c_k,\hat c_k^\dagger]
    \label{eq:weightsmain}
\end{align}
is the commutator of the contribution $\hat c_k=u_k\hat a_k+v_k\hat a_k^\dagger$ of input mode $k$ to $\hat b$. We refer to $w_k$ as the weight of mode $k$ in the output mode. 
For passive transformations, $v_k=0$ and $w_k=|u_k|^2\in[0,1]$ is the fraction of the output mode coming from mode $k$.
Active transformations allow weights outside $[0,1]$, and the normalisation $\sum_kw_k=1$ is then the only constraint.

Given a bipartition $A|B$ of the input modes, the weights of the two subsystems $A$ and $B$ in the output mode are
\begin{align}
    w_A=\sum_{k\in A}w_k ,\qquad w_B=\sum_{k\in B}w_k ,\qquad w_A+w_B=1,
    \label{eq:wAwB}
\end{align}
and they measure how much of the output mode comes from each subsystem. Finally, $\hat\sigma$ denotes the reduced state of the output mode $\hat b$, obtained by tracing out all the other output modes.

We are now ready to state our main result:

\begin{theorem}[PPT inputs and the NW spectrum of an output mode]
\label{thm:single_mode}
Let $\hat\rho$ be an $N$-mode state, $A|B$ a bipartition of its modes, $\hat b$ an output mode of a Gaussian unitary applied to $\hat\rho$, and $\hat\sigma$ the reduced state of this output mode. If $\hat\rho$ is PPT across $A|B$ (in particular if it is separable) then
\begin{align}
    w_A-w_B\in\N(\hat\sigma).
\end{align}
\end{theorem}

Note that nothing is assumed about the input, which may be mixed and non-Gaussian, nor about the Gaussian unitary, which may be passive or active. The theorem is the single-output-mode case of a more general statement, in which several output modes are retained and the NW spectrum is replaced by a positivity condition on their joint characteristic function; this general version is stated and proved in \cref{app:proofmain}.

The value singled out by the theorem will be denoted
\begin{align}
    \hbar_{A|B}(\hat b)\coloneqq w_A-w_B ,
    \label{eq:hbar_AB}
\end{align}
and the reason for this notation is also the idea behind the theorem. 
Partial transposition can be moved from the state to the observable, $\Tr[\hat\rho\hat X]=\Tr[\hat\rho^{\Gamma_B}\hat X^{\Gamma_B}]$ (\cref{sec:background_PPT}), so that the statistics of the output mode $\hat b$ in $\hat\rho$ coincide with those of the operator $\hat b^{\Gamma_B}$ in $\hat\rho^{\Gamma_B}$. 
If $\hat\rho$ is PPT, then $\hat\rho^{\Gamma_B}$ is a quantum state, and this second description is as valid as the first: $\hat\sigma$ also describes the outcomes of measurements of $\hat b^{\Gamma_B}$ on a system prepared in the quantum state $\hat\rho^{\Gamma_B}$.
Transposition exchanges $\hat a_k$ and $\hat a_k^\dagger$: $\hat b^{\Gamma_B}$ is $\hat b$ with $\hat a_k$ and $\hat a_k^\dagger$ exchanged for every mode $k$ of $B$. 
The contributions of $A$ are unchanged, while each contribution $\hat c_k$ of $B$ becomes $u_k\hat a_k^\dagger+v_k\hat a_k$, whose commutator is $-w_k$ instead of $w_k$. 
The weight of $B$ thus changes sign, and
\begin{align}
    \left[\hat b^{\Gamma_B},(\hat b^{\Gamma_B})^\dagger\right]=w_A-w_B=\hbar_{A|B}(\hat b)
\end{align}
instead of $[\hat b,\hat b^\dagger]=w_A+w_B=1$. 
For a PPT input, the statistics of $\hat\sigma$ are thus those of an operator with commutator $\hbar_{A|B}(\hat b)$, rather than $1$, in a valid quantum state. 
Equivalently, $W_{\hat\sigma}$ is the Wigner function of a quantum state for the value $|\hbar_{A|B}(\hat b)|$ of Planck's constant, which is precisely what $\hbar_{A|B}(\hat b)\in\N(\hat\sigma)$ corresponds to (\cref{sec:background_NW}). 

We now comment on a few special cases. 
If the output mode comes from one party only, in which case $w_A\in\{0,1\}$, then $\hbar_{A|B}(\hat b)=\pm1$ and the theorem has no implication, since $\pm1\in\N(\hat\sigma)$ for every state: a mode that does not mix $A$ and $B$ cannot contain meaningful information about their entanglement. 
On the other hand, if the output mode draws equally on both parties, i.e.\ $w_A=w_B=1/2$, then $\hbar_{A|B}(\hat b)=0$ and the theorem implies that $W_{\hat\sigma}\ge0$: in this case, a PPT input cannot produce a Wigner-negative output mode. Apart from these two cases, $|\hbar_{A|B}(\hat b)|<1$ for passive transformations, whereas active transformations can lead to $|\hbar_{A|B}(\hat b)|>1$ (we analyze these two regimes in \cref{sec:entdet}). Finally, exchanging the roles of $A$ and $B$ changes $\hbar_{A|B}(\hat b)$ into its opposite, and since $\N(\hat\sigma)$ is symmetric this gives the same condition, as it should.

\Cref{thm:single_mode} can be used in two ways, illustrated in \cref{fig:detection_generation}.
The first relates to entanglement detection:
taken as a contrapositive, \Cref{thm:single_mode} implies that if $\hbar_{A|B}(\hat b)\notin\N(\hat\sigma)$, then $\hat\rho$ is NPT-entangled across $A|B$. In particular, a single value missing from the NW spectrum of a single output mode witnesses the entanglement of the whole input state. 
The second relates to entanglement generation, and uses the same contrapositive statement with the roles of input and output exchanged: a single-mode state $\hat\sigma$ whose NW spectrum does not contain $\hbar_{A|B}(\hat b)$ cannot be the output mode $\hat b$ of a state that is PPT across $A|B$. 
Hence, if $\hat\sigma$ is sent through $\hat U_G^\dagger$ together with any other modes, in any state whatsoever, the resulting $N$-mode state is NPT-entangled across $A|B$. 
The auxiliary modes play no role here: a gap in the NW spectrum of $\hat\sigma$, which is a form of single-mode nonclassicality, is by itself enough to guarantee entanglement at the output. 
Hereafter, we detail the entanglement detection aspect in \Cref{sec:entdet} and the entanglement generation aspect in
\Cref{sec:entgen}.

\section{Single-mode entanglement detection}
\label{sec:entdet}

\cref{thm:single_mode} provides a general way of detecting NPT entanglement using Gaussian unitaries by measuring a single mode and testing the absence of a specific value in the NW spectrum. In this section, we study in more detail the detection of entanglement via the NW spectrum: in \cref{sec:GaussianNPT}
we consider the case of Gaussian states; in \cref{sec:BSentdet,sec:TMSentdet} we introduce and analyze entanglement detection protocols based on the use of a single beam-splitter (passive) or two-mode squeezer (active), respectively; in \cref{sec:testNW} we explain how to measure the absence of a specific value in the NW spectrum in practice; we finally show in \cref{sec:quantitative} that entanglement witnesses based on the NW spectrum are quantitative, in the sense that they allow us to lower bound the entanglement negativity (see \cref{thm:negativity}).

\subsection{Gaussian NPT entanglement}
\label{sec:GaussianNPT}

We show that the NW spectrum can detect the entanglement of all NPT-entangled Gaussian states. Using the same notations as in \cref{thm:single_mode}:

\begin{theorem}[Gaussian NPT entanglement is detectable with a single mode]
\label{thm:gaussian}
A Gaussian state $\hat\rho$ is NPT across $A|B$ if and only if there exist a
Gaussian unitary and an output mode $\hat b$ such that $2w_A-1\notin\N(\hat\sigma)$.
\end{theorem}

The ``if'' part is \cref{thm:single_mode}. The converse says that, for Gaussian states, the single-mode test is as strong as Simon's criterion: whenever $\hat\rho$ is NPT, one can find an output mode that violates the uncertainty relation it would have to satisfy in the mirror state $\hat\rho^{\Gamma_B}$. The proof, given in \cref{sec:gaussian_case}, is a matter of linear algebra.

\subsection{The beam-splitter protocol}
\label{sec:BSentdet}

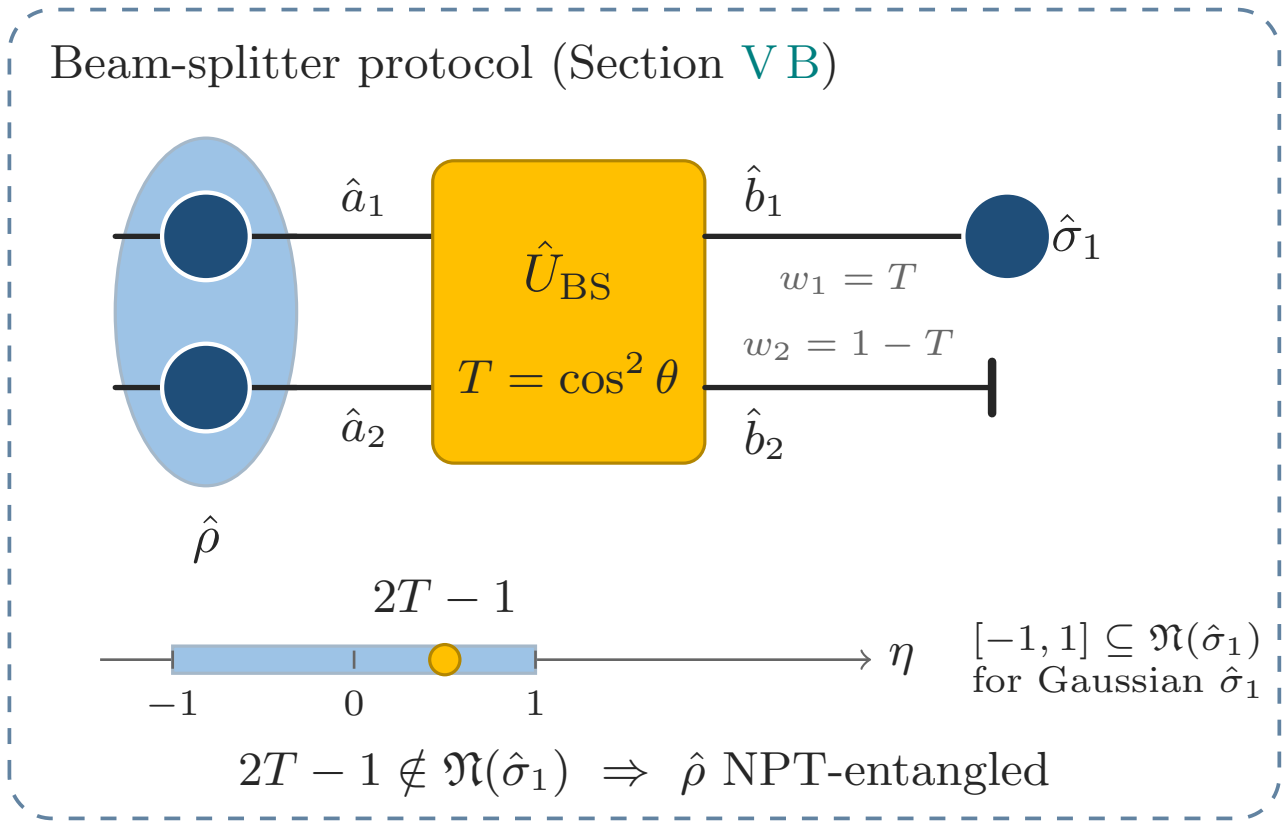
\begin{figure}
    \centering
    \begin{tikzpicture}[font=\normalsize, txt/.style={black!85}]
  \draw[frame] (0.0,-3.35) rectangle (8.4,2.0);
  \node[anchor=north west, txt, font=\small] at (0.15,1.9)
    {Beam-splitter protocol (\cref{sec:BSentdet})};
  \foreach \y in {0.5,-0.5} { \draw[wire] (1.3,\y) -- (2.8,\y); }
  \fill[state] (1.3,0) ellipse (0.6 and 1.15);
  \foreach \y in {0.5,-0.5} {
    \draw[wire] (0.7,\y) -- (1.9,\y);
    \fill[mode] (1.3,\y) circle (0.29);
  }
  \node[txt] at (1.3,-1.5) {$\hat\rho$};
  \node[txt, font=\small, above] at (2.35,0.52) {$\hat a_1$};
  \node[txt, font=\small, below] at (2.35,-0.52) {$\hat a_2$};
  \fill[gate] (2.8,-1.0) rectangle (4.6,1.0);
  \node[txt] at (3.7,0.28) {$\hat U_{\mathrm{BS}}$};
  \node[txt, font=\small] at (3.7,-0.4) {$T=\cos^2\theta$};
  \draw[wire] (4.6,0.5) -- (6.6,0.5);
  \fill[mode] (6.6,0.5) circle (0.29);
  \node[right=5pt, txt] at (6.6,0.5) {$\hat\sigma_1$};
  \node[txt, font=\small, above] at (5.0,0.52) {$\hat b_1$};
  \node[black!60, font=\scriptsize, below=1pt] at (5.55,0.47) {$w_1=T$};
  \draw[wire] (4.6,-0.5) -- (6.5,-0.5);
  \draw[discard] (6.5,-0.67) -- (6.5,-0.33);
  \node[txt, font=\small, below] at (5.0,-0.52) {$\hat b_2$};
  \node[black!60, font=\scriptsize, above=1pt] at (5.55,-0.47) {$w_2=1-T$};
  \draw[black!60, line width=0.5pt, ->] (0.6,-2.3) -- (5.7,-2.3);
  \node[txt, font=\small, right] at (5.7,-2.3) {$\eta$};
  \fill[state] (1.08,-2.39) rectangle (3.48,-2.21);
  \foreach \e/\lab in {-1/$-1$, 0/$0$, 1/$1$} {
    \draw[black!60, line width=0.5pt] ({0.6+1.2*(\e+1.4)},-2.36) -- ({0.6+1.2*(\e+1.4)},-2.24);
    \node[txt, font=\scriptsize, below=1pt] at ({0.6+1.2*(\e+1.4)},-2.36) {\lab};
  }
  \fill[fill=gateorange, draw=gateorange!70!black, line width=0.6pt] (2.88,-2.3) circle (0.1);
  \node[txt, font=\small, above=2pt] at (2.88,-2.2) {$2T-1$};
  \node[txt, font=\scriptsize, anchor=west, align=left] at (6.25,-2.3)
    {$[-1,1]\subseteq\N(\hat\sigma_1)$\\ for Gaussian $\hat\sigma_1$};
  \node[txt, font=\small] at (4.2,-3.05)
    {$2T-1\notin\N(\hat\sigma_1)\ \Rightarrow\ \hat\rho$ NPT-entangled};
\end{tikzpicture}
    \caption{The beam-splitter protocol of \cref{sec:BSentdet}. A two-mode state $\hat\rho$ is sent through a beam-splitter of transmissivity $T=\cos^2\theta$; the output mode $\hat b_1=\cos\theta\,\hat a_1+\sin\theta\,\hat a_2$, with reduced state $\hat\sigma_1$, is retained and the other one is discarded. The weights of the input modes in $\hat b_1$ are $w_1=T$ and $w_2=1-T$, so that \cref{thm:single_mode} tests the value $2T-1$ (orange dot) in $\N(\hat\sigma_1)$. This value lies in $[-1,1]$ (blue), which is contained in the NW spectrum of every Gaussian state: entanglement can only be detected through a non-Gaussian output mode.}
    \label{fig:bs_detection}
\end{figure}

Consider now a generic two-mode bosonic state $\hat{\rho}$. To test entanglement, we can apply \cref{thm:single_mode} choosing a beam-splitter with transmissivity $T = \cos^2 \theta$ as the Gaussian unitary. In this case, if we call the original modes $\hat{a}_1,\hat{a}_2$ and the output modes $\hat{b}_1,\hat{b}_2$, with respect to the $1\vert 2$ partition and choosing $\hat{b}_1 = \cos \theta \  \hat{a}_1 + \sin\theta \ \hat{a}_2 $, we have that $w_{1} = \cos^2 \theta = T$. 
See \cref{fig:bs_detection} for an illustration.
Therefore, we arrive at the following result:
\begin{corollary}[Single-mode entanglement witness with a beam-splitter]
Consider a two-mode state $\hat{\rho}$ passed through a beam-splitter with transmissivity $T$. Call $\hat{\sigma}_1$ the single-mode state of one of the output modes, tracing out the second. If $2 T - 1 \notin \N(\hat{\sigma}_1)$, then $\hat{\rho}$ is entangled. 
\end{corollary}
Notice that choosing $w_2$ would imply that $2 w_2 -1 = 1 - 2T \notin \N (\hat \sigma_1)$, therefore it would amount to test the symmetric point in the NW spectrum, ultimately leading to an equivalent condition. The only additional freedom is in choosing the other single-mode output state after the beam-splitter, since the condition $2T - 1 \notin \N(\hat\sigma_2)$ is generally an independent one. Additionally, since $2T - 1 \in [-1,1]$ and this interval is always filled in the NW spectra of Gaussian states, this protocol can only detect entanglement for non-Gaussian states. 
Finally, let us remark that, by Eq.~\eqref{eq:ncdandNW} with $s = |2T - 1|$, if the nonclassical depth of $\hat \sigma_1$ is strictly greater than $\max\{T, 1-T\}$, then the interval $[-|2T - 1|, |2T -1|]$ is empty in $\N(\hat\sigma_1)$, therefore single-mode nonclassicality as quantified by the nonclassical depth can be used to certify two-mode entanglement.

The beam-splitter protocol extends to $N$ modes by sending a multimode state $\hat{\rho}$ through a network of beam-splitters and keeping a single output mode $\hat{b} = \sum_k u_k \hat{a}_k$, with weights $w_k = |u_k|^2$ determined by the transmissivities of each beam-splitter. If $\hat{\rho}$ is fully separable, or merely PPT across every bipartition, \cref{thm:single_mode} and the symmetry of the NW spectrum give $\pm(2 w_A - 1) \in \N(\hat{\sigma})$ for every subset $A$ of the input modes, where $w_A = \sum_{k \in A} w_k$: each missing value certifies NPT entanglement across the corresponding bipartition. Genuine multipartite entanglement (GME) requires more, since a biseparable state $\hat{\rho} = \sum_{A} p_A \hat{\rho}_{A}$, with $\hat{\rho}_A$ separable across $A|\bar{A}$, only yields $\hat{\sigma} = \sum_A p_A \hat{\sigma}_A$ with $\pm(2w_A - 1) \in \N(\hat{\sigma}_A)$ for each term, and the convex mixture property retains only the values common to all terms. If the transmissivities are chosen such that $|2 w_A - 1| = c$ for all bipartitions, then $c \in \N(\hat{\sigma})$ for every biseparable, and more generally every PPT-mixture \cite{jungnitsch2011taming}, input, so that $c \notin \N(\hat{\sigma})$ certifies GME. This is possible only for $N = 3$, with $w_k = 1/3$ and $c = 1/3$, i.e.\ for one output mode of a balanced tritter: for $N \geq 4$, equal values of $|2w_A - 1|$ for all single modes and pairs would force $w_k = 1/3$ for all $k$. In general, all the values $\pm(2w_A - 1)$ lie in $[-s, s]$ with $s = \max_A |2w_A - 1|$, and the argument of \cref{sec:ncdandNW} applied to each term shows that $W^{(-s)}_{\hat{\sigma}} \geq 0$ for every PPT-mixture input, i.e.\ $\tau_{\hat{\sigma}} \leq (1+s)/2$. For equal weights $w_k = 1/N$, which minimise $s$ among passive networks, $s = 1 - 2/N$, so that $\tau_{\hat{\sigma}} > 1 - 1/N$ certifies GME: this is the criterion of Ref.~\cite{zaw2026sufficient}, which we recover here with the biseparable set enlarged to PPT mixtures.

\subsection{The two-mode squeezing protocol}
\label{sec:TMSentdet}

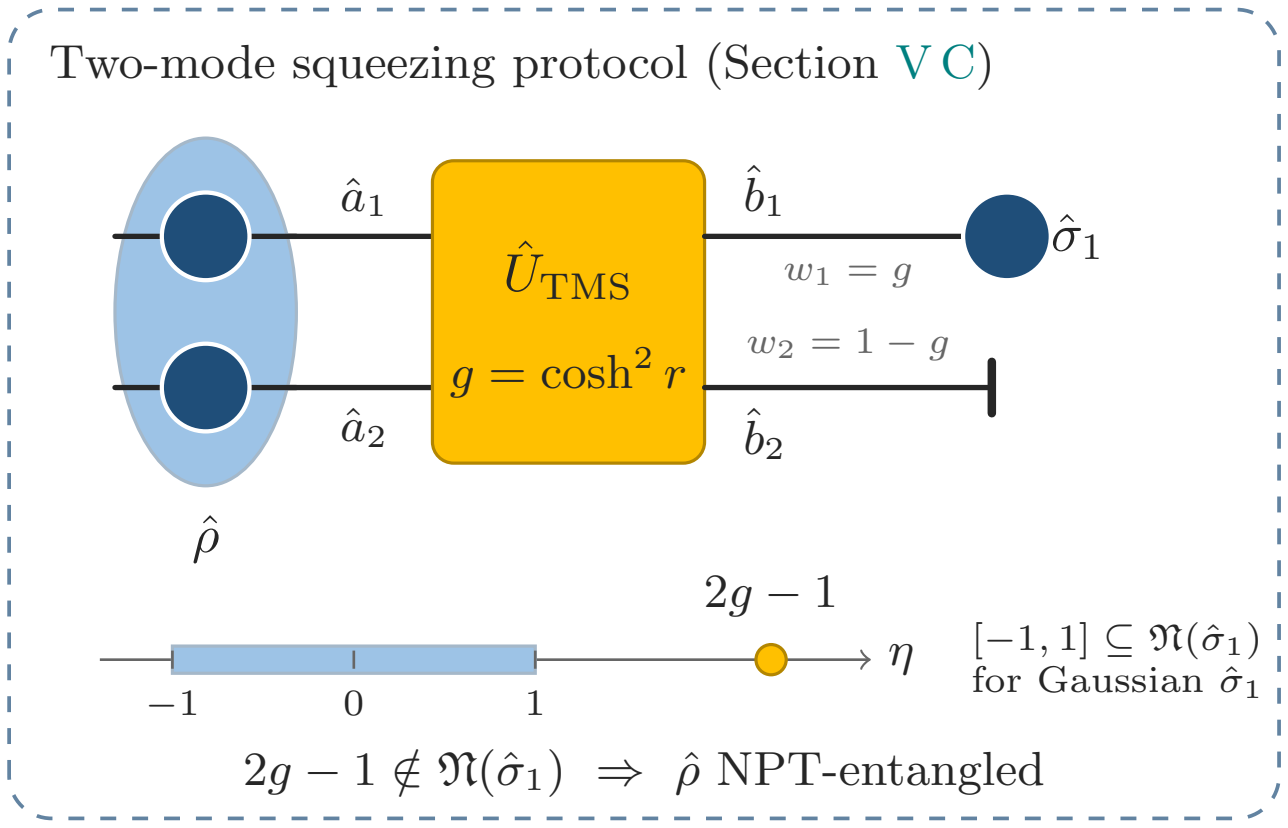
\begin{figure}
    \centering
    \begin{tikzpicture}[font=\normalsize, txt/.style={black!85}]
  \draw[frame] (0.0,-3.35) rectangle (8.4,2.0);
  \node[anchor=north west, txt, font=\small] at (0.15,1.9)
    {Two-mode squeezing protocol (\cref{sec:TMSentdet})};
  \foreach \y in {0.5,-0.5} { \draw[wire] (1.3,\y) -- (2.8,\y); }
  \fill[state] (1.3,0) ellipse (0.6 and 1.15);
  \foreach \y in {0.5,-0.5} {
    \draw[wire] (0.7,\y) -- (1.9,\y);
    \fill[mode] (1.3,\y) circle (0.29);
  }
  \node[txt] at (1.3,-1.5) {$\hat\rho$};
  \node[txt, font=\small, above] at (2.35,0.52) {$\hat a_1$};
  \node[txt, font=\small, below] at (2.35,-0.52) {$\hat a_2$};
  \fill[gate] (2.8,-1.0) rectangle (4.6,1.0);
  \node[txt] at (3.7,0.28) {$\hat U_{\mathrm{TMS}}$};
  \node[txt, font=\small] at (3.7,-0.4) {$g=\cosh^2 r$};
  \draw[wire] (4.6,0.5) -- (6.6,0.5);
  \fill[mode] (6.6,0.5) circle (0.29);
  \node[right=5pt, txt] at (6.6,0.5) {$\hat\sigma_1$};
  \node[txt, font=\small, above] at (5.0,0.52) {$\hat b_1$};
  \node[black!60, font=\scriptsize, below=1pt] at (5.55,0.47) {$w_1=g$};
  \draw[wire] (4.6,-0.5) -- (6.5,-0.5);
  \draw[discard] (6.5,-0.67) -- (6.5,-0.33);
  \node[txt, font=\small, below] at (5.0,-0.52) {$\hat b_2$};
  \node[black!60, font=\scriptsize, above=1pt] at (5.55,-0.47) {$w_2=1-g$};
  \draw[black!60, line width=0.5pt, ->] (0.6,-2.3) -- (5.7,-2.3);
  \node[txt, font=\small, right] at (5.7,-2.3) {$\eta$};
  \fill[state] (1.08,-2.39) rectangle (3.48,-2.21);
  \foreach \e/\lab in {-1/$-1$, 0/$0$, 1/$1$} {
    \draw[black!60, line width=0.5pt] ({0.6+1.2*(\e+1.4)},-2.36) -- ({0.6+1.2*(\e+1.4)},-2.24);
    \node[txt, font=\scriptsize, below=1pt] at ({0.6+1.2*(\e+1.4)},-2.36) {\lab};
  }
  \fill[fill=gateorange, draw=gateorange!70!black, line width=0.6pt] (5.04,-2.3) circle (0.1);
  \node[txt, font=\small, above=2pt] at (5.04,-2.2) {$2g-1$};
  \node[txt, font=\scriptsize, anchor=west, align=left] at (6.25,-2.3)
    {$[-1,1]\subseteq\N(\hat\sigma_1)$\\ for Gaussian $\hat\sigma_1$};
  \node[txt, font=\small] at (4.2,-3.05)
    {$2g-1\notin\N(\hat\sigma_1)\ \Rightarrow\ \hat\rho$ NPT-entangled};
\end{tikzpicture}
    \caption{The two-mode squeezing protocol of \cref{sec:TMSentdet}. The beam-splitter is replaced by a two-mode squeezer of gain $g=\cosh^2 r$, with output mode $\hat b_1=\cosh r\,\hat a_1+\sinh r\,\hat a_2^\dagger$. The weights are now $w_1=g$ and $w_2=1-g$, and the tested value $2g-1=\cosh 2r$ lies outside $[-1,1]$, so that Gaussian entanglement is within reach: for a Gaussian output mode with symplectic eigenvalue $\nu$, a gain $g>\nu+1/2$ suffices.}
    \label{fig:tms_detection}
\end{figure}

Consider again a two-mode state $\hat{\rho}$, and choose now as the Gaussian unitary a two-mode squeezer $\hat{U}_{\mathrm{TMS}}(r) = \exp[r(\hat{a}_1^\dagger \hat{a}_2^\dagger - \hat{a}_1 \hat{a}_2)]$ with squeezing parameter $r>0$, or equivalently gain $g = \cosh^2 r > 1$. The output modes are $\hat{b}_1 = \cosh r \ \hat{a}_1 + \sinh r \ \hat{a}_2^\dagger$ and $\hat{b}_2 = \cosh r \ \hat{a}_2 + \sinh r \ \hat{a}_1^\dagger$, so that, with respect to the $1 \vert 2$ partition and choosing $\hat{b}_1$, we have $w_1 = \cosh^2 r = g$ and $2 w_1 - 1 = 2g - 1$. 
See \cref{fig:tms_detection} for an illustration.
\cref{thm:single_mode} then leads to the following result:
\begin{corollary}[Single-mode entanglement witness with a two-mode squeezer]\label{coro:tms}
Consider a two-mode state $\hat{\rho}$ passed through a two-mode squeezer with gain $g>1$. Call $\hat{\sigma}_1$ the single-mode state of one of the output modes, tracing out the second. If $2g - 1 \notin \N(\hat{\sigma}_1)$, then $\hat{\rho}$ is entangled.
\end{corollary}
As for the beam-splitter, choosing $w_2 = -\sinh^2 r = 1-g$ would amount to test the symmetric point $1 - 2g$, hence the same condition, while the second output mode provides an independent one. The crucial difference with the beam-splitter protocol is that $2g - 1 > 1$ lies outside the interval $[-1,1]$, so that Gaussian entanglement is no longer out of reach. Indeed, if $\eta \in \N(\hat{\sigma}_1)$ then, by definition of the NW spectrum, $W_{\hat{\sigma}_1}$ is the Wigner function of a quantum state once $\eta$ replaces $\hbar$ in the Wigner transform (see \cref{app:NWproperties,app:gabor}): its covariance matrix must then satisfy $\bm{V}_{\hat{\sigma}_1} + \frac{i\eta}{2}\bm{\Omega} \succeq 0$ \cite{Narcowich1986}, i.e.\ $|\eta| \leq 2\nu$ with $\nu$ the symplectic eigenvalue of $\bm{V}_{\hat{\sigma}_1}$, and its purity $2\pi\eta \int W_{\hat{\sigma}_1}^2 = \eta \, \Tr[\hat{\sigma}_1^2]$ cannot exceed one, i.e.\ $|\eta| \leq 1/\mu$ with $\mu = \Tr[\hat{\sigma}_1^2]$. Therefore
\begin{align}
    \N(\hat{\sigma}_1) \subseteq [-2\nu, 2\nu] \cap [-1/\mu, 1/\mu]
    \label{eq:NWbounds}
\end{align}
for any single-mode state, and $\N(\hat{\sigma}_1) = [-2\nu, 2\nu]$ for Gaussian states. A gain $g > \nu + 1/2$, or $g > (1 + 1/\mu)/2$, is thus sufficient to certify the entanglement of $\hat{\rho}$ from the covariance matrix or the purity of the output mode alone; in particular, a pure output mode certifies entanglement for any $g > 1$. For a Gaussian output mode the condition $g > \nu + 1/2$ is also necessary, consistently with \cref{thm:gaussian}. 

Finally, let us remark that by the Bloch--Messiah decomposition $\hat{U}_{\mathrm{TMS}}(r) = \hat{U}_{\mathrm{BS}}[ \hat{S}(r) \otimes \hat{S}(-r)] \hat{U}_{\mathrm{BS}}^\dagger$, with $\hat{U}_{\mathrm{BS}}$ a balanced beam-splitter and $\hat{S}(\pm r)$ single-mode squeezers, the active element of the protocol can be moved to the measurement stage: the single-mode squeezers and final beam-splitter can be absorbed into unbalanced heterodyne detections and classical post-processing~\cite{chabaud2020certification}, leaving only passive optics before the detectors.

\subsection{Measuring the NW spectrum}
\label{sec:testNW}

One can test the KLM conditions through which the NW spectrum is implicitly defined (see \cref{sec:background_NW}). A major obstacle to this end is the arbitrariness of the choice of phase-space points: in principle, for any natural number $N>1$, one would have to check all possible choices of $N$ points in the phase-space of the interested mode(s) until a KLM matrix with a negative eigenvalue is found. 

A way around this obstacle was found by Cordero, de Gosson and Nicola \cite{cordero2019}, who showed that the arbitrary families of phase-space points can be replaced by the points of a fixed lattice $\Lambda \subset \reals^2$, at the price of constructing the KLM matrix with respect to a fiducial pure state $\vert \phi \rangle$, the \emph{window}. For $\lambda = (\lambda_x, \lambda_p) \in \Lambda$ let $\phi_\lambda = \hat{T}_\eta(\lambda) \phi$, where $\hat{T}_\eta(\lambda) = \hat{D}(\lambda_x, \lambda_p/\eta)$ is the displacement operator to the phase-space point $(\lambda_x, \lambda_p/\eta)$ and $\phi: \mathbb{R} \to \mathbb{C}$ is the wavefunction of the fiducial pure state, and let $\hat{A}_\eta[\hat{\sigma}]$ be the operator whose Wigner function is $\eta \, W_{\hat{\sigma}}(x, \eta p)$, so that $\eta \in \N(\hat{\sigma})$ if and only if $\hat{A}_\eta[\hat{\sigma}] \succeq 0$ (\cref{app:NWproperties,app:gabor}). If the family $\{\phi_\lambda\}_{\lambda \in \Lambda}$ is a Gabor frame, i.e.\ a (generally overcomplete) family spanning the whole Hilbert space in a stable way, then $\eta \in \N(\hat{\sigma})$ if and only if
\begin{align}
    \bm{K}_F = \Big[ \bra{\phi_\lambda} \hat{A}_\eta[\hat{\sigma}] \ket{\phi_{\mu}} \Big]_{\lambda, \mu \in F} \succeq 0
    \label{eq:gabor_test}
\end{align}
for every finite subset $F \subset \Lambda$. In other words, $\bm{K}_F$ is the matrix of the quadratic form of $\hat{A}_\eta[\hat{\sigma}]$ on a finite subset of displaced fiducial states, and the frame property ensures that these are dense in the Hilbert space (\cref{app:gabor}). The $\phi_\lambda$ are thus copies of the window wavefunction along the lattice with spacings rescaled by $1/\eta$ along $p$; for the vacuum window they are coherent states. For a Gaussian window of any width and $\Lambda = a\mathbb{Z} \times b\mathbb{Z}$, the frame property holds if and only if $ab < 2\pi\eta$ \cite{lyubarskii1992,seip1992}.

In practice, Eq.~\eqref{eq:gabor_test} provides a hierarchy of finite tests: taking for $F$ the $(2n+1)^2$ lattice points closest to the origin, $n = 1, 2, \dots$, each level is a necessary condition for $\eta \in \N(\hat{\sigma})$, and a negative eigenvalue of $\bm{K}_F$ at any level certifies that $\eta \notin \N(\hat{\sigma})$, hence the entanglement of the input state. To fill in these matrices through data that can be directly measured on $\hat{\sigma}$, we show in \cref{app:gabor} that $\bra{\phi_\lambda} \hat{A}_\eta[\hat{\sigma}] \ket{\phi_\mu} = \Tr[\hat{\sigma} \hat{O}_{\lambda\mu}]$, where $\hat{O}_{\lambda\mu}$ is the operator whose Wigner function is that of $\ket{\phi_\mu}\!\bra{\phi_\lambda}$ stretched by a factor $\eta$ along $p$. To get these values from Gaussian measurements, we take the fiducial wavefunction to be Gaussian to start with. 

In that case, for $\eta > 1$, which is the case of interest for the two-mode squeezing protocol, $\hat{O}_{\lambda\lambda}/\eta$ is a mixed Gaussian state, and choosing for the window the squeezed vacuum with variance $\eta/2$ along $x$ it is the thermal state with mean photon number $(\eta-1)/2$ displaced to $\lambda$. Likewise, off-diagonal operators are also Gaussian mixtures of coherent states, and their expectation values are sample means of explicit bounded functions of the outcomes of heterodyne detection, Eq.~\eqref{eq:app_g}:
\begin{align}
    \bra{\phi_\lambda} \hat{A}_\eta[\hat{\sigma}] \ket{\phi_{\mu}} = \int \mathrm{d}^2 z \ Q_{\hat{\sigma}}(z) \, g_{\lambda \mu}(z),
    \label{eq:gabor_heterodyne}
\end{align}
with $Q_{\hat{\sigma}}$ the Husimi function of $\hat{\sigma}$. A single set of heterodyne data thus fills all the entries of $\bm{K}_F$ at all levels of the hierarchy, without bias.

For $\eta < 1$, however, the operators $\hat{O}_{\lambda\lambda}$ violate the uncertainty relations for Gaussian states (they are not positive-semidefinite operators). In particular, their $P$-function is not a regular non-negative function, so that no window choice allows to estimate their expectation values directly through heterodyne detection (balanced or not); they can instead be estimated from phase-randomised homodyne detection, since their characteristic functions are Gaussian \cite{lvovsky2009continuous}, or from heterodyne detection at the price of a regularisation bias \cite{chabaud2020building,chabaud2020certification}.

\subsection{Quantitative bounds}
\label{sec:quantitative}

The NW spectrum is tested through the positivity of KLM matrices: $\eta \notin \N(\hat{\sigma})$ means that, for some choice of points $x_1, \dots, x_m$ in the phase space of the output mode, the matrix
\begin{align}
    \bm{M}_{jk} = \chi_{\hat{\sigma}}(x_j - x_k) \, e^{\frac{i\eta}{2} x_j^T \bm{\Omega} x_k}, \qquad j,k = 1, \dots, m,
    \label{eq:KLMmatrix}
\end{align}
has a negative eigenvalue. When this happens for the value $\eta = 2w_A - 1$ prescribed by \cref{thm:single_mode}, the negative eigenvalues do not only certify entanglement, they also quantify it. Let $N(\hat{\rho}) = (\| \hat{\rho}^{\Gamma_B} \|_1 - 1)/2$ and $E_N(\hat{\rho}) = \log_2 \| \hat{\rho}^{\Gamma_B} \|_1$ be the negativity and the logarithmic negativity of $\hat{\rho}$ across $A|B$ \cite{vidal2002computable}, and let $\bm{M}_-$ be the negative part of $\bm{M}$, so that $\Tr \bm{M}_-$ is the sum of the moduli of its negative eigenvalues.

\begin{theorem}[Negativity bound]
\label{thm:negativity}
With the notation of \cref{thm:single_mode}, let $\eta = 2w_A - 1$ and let $\bm{M}$ be the KLM matrix \eqref{eq:KLMmatrix} of $\hat{\sigma}$ for any $m$ points. Then
\begin{equation} 
\begin{aligned}
    & N(\hat{\rho}) \geq \frac{\Tr \bm{M}_-}{m} \geq \frac{\max\{0, -\lambda_{\min}(\bm{M})\}}{m},
    \\
    & E_N(\hat{\rho}) \geq \log_2 \frac{\| \bm{M} \|_1}{m}.
    \label{eq:negativity_bound}
\end{aligned}
\end{equation}
\end{theorem}

We refer to \cref{app:quantitative} for a proof, which also gives the sharper bound $N(\hat{\rho}) \geq -\lambda_{\min}(\bm{M}) / \| v \|_1^2$ in terms of the normalised eigenvector $v$ of the smallest eigenvalue. The division by $m$ is what makes the bound meaningful: $\bm{M}$ has unit diagonal, so that $\bm{M}/m$ has unit trace and plays the role of a density matrix whose negative eigenvalues are compared with those of $\hat{\rho}^{\Gamma_B}$. The same argument applies to the matrices $\bm{K}_F$ of \cref{sec:testNW}, with $m$ replaced by the largest eigenvalue of the Gram matrix $[\braket{\phi_\lambda | \phi_\mu}]_{\lambda, \mu \in F}$ of the displaced windows. Let us remark that the bound requires the value of $\eta$ prescribed by the Gaussian unitary and by the bipartition: the absence of any other value from $\N(\hat{\sigma})$ carries no information on the entanglement of $\hat{\rho}$. On the other hand, it does not require any Wigner negativity, nor the input to be non-Gaussian. For instance, a two-mode squeezed vacuum with squeezing parameter $r$ sent through the inverse two-mode squeezer gives the vacuum as output, with $\eta = \cosh 2r$; the three points $x_1 = (0,0)$, $x_2 = (s,0)$ and $x_3 = (0,s)$ already give a negative eigenvalue for small enough $s$ (\cref{app:quantitative}), and for $r = 1/2$ and $s = 1$ the bound reads $N(\hat{\rho}) \geq 0.0117$, the exact value being $(e^{2r} - 1)/2 \approx 0.859$.

\Cref{thm:negativity} belongs to a line of works which turn Bochner-type conditions on the characteristic function into quantitative witnesses. 
Ref.~\cite{mari2011directly} used the matrices~\eqref{eq:KLMmatrix} with $\eta=0$ to certify the Wigner negativity of a single mode from a few measured values of its characteristic function, their negative eigenvalues bounding the distance of the state to the set of Wigner-positive states. 
Ref.~\cite{zaw2024certifiable} applied the same matrices to the output of a balanced beam-splitter, where Wigner negativity signals entanglement, and obtained lower bounds on the trace distance of the input to the separable states. 
Ref.~\cite{zaw2026sufficient} did the same for balanced beam-splitter networks and the distance to the biseparable states, thus quantifying genuine multipartite entanglement. 
\Cref{thm:negativity} takes one more step. 
Using in Eq.~\eqref{eq:KLMmatrix} the value $\eta=2w_A-1$ prescribed by the Gaussian unitary and the bipartition, it applies to arbitrary Gaussian unitaries, in particular unbalanced and active ones, hence also to Gaussian entanglement. 
Moreover, it bounds a standard entanglement monotone, the negativity, rather than a distance.

\section{Single-mode entanglement generation}
\label{sec:entgen}

Beyond entanglement detection, another interesting consequence of \cref{thm:single_mode} pertains to sufficient conditions for entanglement generation via a Gaussian unitary. Consider an $N$-mode state $\hat\sigma$ as well as a Gaussian unitary $\hat U_G$ with symplectic matrix $\bm S$. Given any bipartition $A|B$ of the $N$ output modes, we obtain a simple condition on $\hat\sigma$ which ensures that the state $\hat U_G^\dag\hat\sigma\hat U_G$ is entangled over $A|B$:

\begin{corollary}\label{coro:entgen}
    Let $\hat\sigma_1$ denote the reduced state over the first mode of $\hat\sigma$. If $2w_A-1\notin\N(\hat\sigma_1)$, then the state $\hat U_G^\dag\hat\sigma\hat U_G$ is NPT-entangled over $A|B$.
\end{corollary}

\begin{proof} Here, the first mode of $\hat\sigma$ plays the role of the mode $\hat b$ in \cref{thm:single_mode}. Let $\hat\rho=\hat U_G^\dag\hat\sigma\hat U_G$ and assume that the state $\hat\rho=\hat U_G^\dag\hat\sigma\hat U_G$ is not NPT-entangled across $A|B$, i.e.\ it is PPT across $A|B$. Then by \cref{thm:single_mode}, the first mode of $\hat U_G\hat\rho\hat U_G^\dag$ has the value $2w_A-1$ in its NW spectrum. But $\hat U_G\hat\rho\hat U_G^\dag=\hat\sigma$, so its first mode is precisely $\hat\sigma_1$ and thus $2w_A-1\in\N(\hat\sigma_1)$. Hence, $2w_A-1\notin\N(\hat\sigma_1)$ implies that $\hat U_G^\dag\hat\sigma\hat U_G$ is NPT-entangled across $A|B$.
\end{proof}

When restricting to two modes, \cref{coro:entgen} has simple consequences which are worth pointing out. First, if the Gaussian unitary is a two-mode squeezing unitary, we obtain the converse of \cref{thm:gaussian,coro:tms}: high enough squeezing gain entangles any tensor product of Gaussian states, in contrast to beam-splitters. Second, when the Gaussian unitary is a two-mode beam-splitter unitary, \cref{coro:entgen} provides a sufficient condition for entanglement generation which subsumes and generalizes that of \cite{gholipour2016entanglement} based on nonclassical depth due to \cref{eq:ncdandNW}. Moreover, it also shows that (unbalanced) beam-splitter entanglement due to nonclassical depth is always NPT, a fact that was only proven recently in the case of balanced beam-splitter~\cite{zaw2024certifiable}.

To further illustrate the application of \cref{coro:entgen} to beam-splitter interactions, consider injecting the state $\hat{\rho}_q$, whose NW spectrum is given in \cref{1PSSTS}, into one input port of a beam-splitter of transmissivity $T$, while leaving the other input in an arbitrary state. 
According to \cref{coro:entgen}, NPT entanglement is guaranteed for every such input whenever the tested value $2T-1$ lies outside the NW spectrum of $\hat{\rho}_q$, i.e., by \cref{eq:probabilistic_NW_interval}, whenever $|2T-1|<\eta_-$ or $|2T-1|>\eta_+$. Since $\eta_+\geq1$, the second condition is never met, and the condition reads
\begin{equation}
\frac{1-\eta_-}{2}<T<\frac{1+\eta_-}{2} .
\label{eq:probabilistic_entanglement_interval}
\end{equation}
The nonclassical depth alone, through \cref{eq:ncdandNW}, only guarantees this for $|2T-1|<s_{\max}$: since $\eta_-\geq s_{\max}$, with equality only for $q=0$, the NW-spectrum criterion captures a strictly larger range of transmissivities for $0<q<q_{\max}$, irrespective of the state injected into the second input port.

Consider also the second example, state $\hat{\rho}_{\mathrm{2PSSV}}(p)$ in \cref{eq:2PSSV}. For $p \simeq 0.45$, its NW spectrum intersects the interval $(-1,1)$ along two symmetric segments, which are separated from both the extremes and the centre $\eta = 0$. Therefore, with these states, we will have the following peculiar behavior: mixing them at a beam-splitter with any other single-mode state, NPT entanglement will be generated for arbitrarily small values of the mixing angle up to a certain value $\cos^2 \theta_{+} = (1+\eta_{+})/2$, marking the upper end of the segment belonging to the NW spectrum and located above $0$ and below $1$. From there on, increasing $\theta$~\footnote{And therefore moving from right to left in the NW spectrum, as the tested value $2T-1=2\cos^2 \theta-1$ decreases} for a continuous range of mixing angles, we would not be able to certify NPT entanglement generation for arbitrary inputs in the second mode, up to a second threshold $\cos^2 \theta_{-} = (1+\eta_{-})/2$, marking the end of the segment belonging to the NW spectrum. Increasing further the mixing angle, NPT entanglement will become again certifiable, up and past the balanced BS angle, after which the configurations repeat symmetrically by exchanging the labels of the output modes. This showcases, in particular, a whole range of mixing angles between $0$ and $\theta_{+}$ for which NPT entanglement is generated.

\section{Discussion and outlook}
\label{sec:concl}

In this work, we have strengthened the existing connection between nonclassicality and entanglement~\cite{kim2002entanglement,wang2002theorem,asboth2005computable,vogel2014unified,gholipour2016entanglement} using the NW spectrum. We obtained simple sufficient conditions which ensure that the output of a Gaussian unitary evolution is entangled based on the NW spectrum of any of its input modes. Conversely, these conditions imply that measuring the NW spectrum of a single mode after a Gaussian unitary can quantitatively witness the NPT entanglement structure of multimode quantum states, and in particular of any Gaussian state, and even GME of specific states. As an entanglement detection tool, the NW spectrum provides strictly stronger entanglement witnesses than previous work based on nonclassical depth~\cite{gholipour2016entanglement}.

Our work opens several interesting directions. The NW spectrum can serve as an entanglement witness and we provide a prescription for how to measure the NW spectrum of a single-mode state using homodyne or heterodyne detection. In light of a potential experimental demonstration, it would be valuable to assess the sample complexity of this method, as well as its robustness to noise, losses and imperfect detection. On the more theoretical side, we have tightened the sufficient conditions for single-mode nonclassicality to induce entanglement via beam-splitters with arbitrary auxiliary state \cite{gholipour2016entanglement}. While it is known that optical nonclassicality is necessary~\cite{kim2002entanglement,wang2002theorem}, it would be desirable to obtain conditions that are both necessary and sufficient. Existing examples of highly nonclassical states that do not induce beam-splitter entanglement~\cite{goldberg2018nonclassical} suggest that the answer to this problem may be quite subtle. Finally, it would be interesting to investigate further the connection between the NW spectrum and GME as well as non-Gaussian entanglement~\cite{zaw2024certifiable,zaw2026sufficient}.
More generally, our work indicates that the NW spectrum is a very powerful tool for investigating fundamental properties of photonic quantum states in phase-space. As such, we believe that it could find further applications in quantum information theory.

\section*{Acknowledgements}

UC acknowledges interesting discussions with KN Huynh-Vu. MF and MW acknowledge fruitful discussions with Elizabeth Agudelo and Antoine Debray.
ZVH, JD, HH, MW and UC acknowledge funding from the European Union’s Horizon Europe Framework Programme (EIC Pathfinder Challenge project Veriqub) under Grant Agreement No.~101114899. MF and MW acknowledge that this project has received funding from the European Union’s Horizon Europe research and innovation programme under the Marie Skłodowska-Curie grant agreement No.~101208849. NCD and JNP acknowledge funding from Funda\c{c}\~ao para a Ci\^encia e Tecnologia (FCT) via the research center GFM project UID/00208/2025.

\section*{AI disclosure}

The authors acknowledge the use of AI tools to assist with manuscript preparation and the
development and verification of mathematical proofs.
The research question and approach were fully developed by the authors, who take full responsibility for the results presented in the manuscript.

\bibliography{bib,biblio}

\widetext

\appendix
\crefalias{section}{appendix}

\section{Properties of the NW spectrum}
\label{app:NWproperties}

This section collects the main properties of the NW
spectrum that are relevant for the present work.

\subsection{Main properties}

We use the conventions fixed in the main text for
the symplectic Fourier transform and for functions of $\eta$-positive
type.  The symmetry, compactness, and convolution properties below are
standard; see \cite[Proposition~1.2]{diasprata2009} and the original
papers \cite{narcowich_conditions_1988,Narcowich1986}.  

\begin{theorem}[Basic properties of the NW spectrum]
\label{TheoremPropertiesNW1}
Let $F,G,F_1,F_2 \in L^2(\mathbb{R}^{2N})$ be nonzero real functions whose symplectic
Fourier transforms are continuous.  Whenever a convolution is used, assume
in addition that it is well defined (for instance, $F,G,F_1,F_2\in L^1\cap L^2$).
Then:
\begin{enumerate}
\item {\bf Symmetry:}
\begin{equation}
  \eta\in\N(F) \quad\Longleftrightarrow\quad -\eta\in\N(F).
\end{equation}

\item {\bf Compactness:} $\N(F)$ is a compact subset of $\mathbb{R}$.

\item {\bf Convolution:}
\begin{equation}
  \N(F)+\N(G)\subseteq\N(F*G),
  \label{eq:nw_prop_convolution}
\end{equation}
where $*$ denotes convolution, and the sum on the left is the Minkowski sum.

\item {\bf Convex mixtures:} Let $F=\sum_i p_iF_i$, where $p_i\geq0$,
$\sum_i p_i=1$. Then
\begin{equation}
  \bigcap_i\N(F_i)\subseteq\N(F).
\end{equation}
The same conclusion holds for probability mixtures
$F=\int F_s\,d\mu(s)$ assuming convergence.

\item {\bf Rescaling property I:} Let $\lambda\neq0$ and consider the rescaling $ F_\lambda(z)=\lambda^{-2n}F( z/\lambda)$.
We have:
\begin{equation}
  \eta\in\N(F_\lambda)
  \quad\Longleftrightarrow\quad
  \lambda^{-2}\eta\in\N(F),
  \qquad\text{hence}\qquad
  \N(F_\lambda)=\lambda^2\N(F).
\end{equation}

\item {\bf Tensor products:} let $F_j$, $j=1,2$ 
functions on $\mathbb{R}^{2N_j}$.
Then
\begin{equation}
  \N(F_1\otimes F_2)=\N(F_1)\cap\N(F_2).
\end{equation}

\item {\bf Traces:} let $F$ be a normalized phase-space function on
$\mathbb{R}^{2N_1}\times\mathbb{R}^{2N_2}$, and let $F_1(z_1)=\int_{\mathbb{R}^{2N_2}}F(z_1,z_2) d z_2$, $F_2(z_2)=\int_{\mathbb{R}^{2N_1}}F(z_1,z_2) d z_1$ be the reduced functions.  Then
\begin{equation}
  \N(F)\subseteq\N(F_1)\cap\N(F_2).
\end{equation}
In particular, for a bipartite state $\rho$ with reduced states
$\rho_1=\Tr_2\rho$ and $\rho_2=\Tr_1\rho$,
\begin{equation}
  \N(W_\rho)\subseteq
  \N(W_{\rho_1})\cap\N(W_{\rho_2}).
\end{equation}

\item {\bf Rescaling property II:} For $\lambda>0$ define the trace-preserving ``super’’ operator $\Theta_\lambda$ weakly by
\begin{equation}
 \chi_{\Theta_\lambda[\rho]}(\xi)=\chi_\rho(\lambda\xi).
\end{equation}
where $\chi$ is the characteristic function of the state $\rho$. Notice that this is tantamount to the rescaling in 5. We then have
\begin{equation}
 \Theta_\lambda[\rho]\geq0
 \quad\Longleftrightarrow\quad
 \lambda^{-2}\in\N(W_\rho).
\end{equation}
Moreover, $\Theta_\lambda$ commutes with partial traces.  Consequently,
for a bipartite state $\rho$,
\begin{equation}
 \Theta_\lambda[\rho]\geq0
 \quad\Longrightarrow\quad
 \Theta_\lambda[\rho_1]\geq0
 \ \text{and}\
 \Theta_\lambda[\rho_2]\geq0,
\end{equation}
which is equivalent to the marginal inclusion above.  If $\rho=\rho_1\otimes\rho_2$ is a product
state, the implication is an equivalence.

\end{enumerate}
\end{theorem}

\begin{proof}
For a finite set $a_1,\dots,a_n$ in phase space, write $\bm M(\eta)$ more explicitly as
\begin{equation}
 \bm M_F^{(\eta)}(a_1,\dots,a_n)
 =\left[
 \widetilde F(a_j-a_k)
 \exp\!\left(\frac{i\eta}{2} a^T_j \bm\Omega \,a_k \right)
 \right]_{j,k=1}^n .
\end{equation}
By the KLM criterion, $\eta\in\N(F)$ precisely when all these matrices
are positive semidefinite.

\smallskip
\noindent\emph{(1) Symmetry.}
Since $F$ is real,
$\widetilde F(-a)=\overline{\widetilde F(a)}$.  Consequently,
$\bm M_F^{(-\eta)}=\overline{\bm M_F^{(\eta)}}$ (up to the simultaneous
replacement $a_j\mapsto -a_j$).  A Hermitian matrix is positive
semidefinite if and only if its complex conjugate is positive
semidefinite.  This proves the claim.

\smallskip
\noindent\emph{(2) Compactness.}
Closedness follows directly from continuity in $\eta$: if
$\eta_m\to\eta$ and every $\bm M_F^{(\eta_m)}$ is positive semidefinite, then
$\bm M_F^{(\eta_m)}\to \bm M_F^{(\eta)}$, and the cone of positive semidefinite
matrices is closed.  Boundedness follows from the
boundedness theorem for spectra of continuous functions of
$\eta$-positive type; see \cite[Proposition~1.2(i)]{diasprata2009} or
\cite{narcowich_conditions_1988}.  Hence $\N(F)$ is closed and bounded, and therefore
compact.

\smallskip
\noindent\emph{(3) Convolution.}
Take $\alpha\in\N(F)$ and $\beta\in\N(G)$.  Since
$\widetilde{F*G}=\widetilde F\,\widetilde G$ and
\begin{equation}
 e^{\frac{i(\alpha+\beta)}2 a^T_j \bm\Omega \,a_k}
 =e^{\frac{i\alpha}2 a^T_j \bm\Omega \,a_k}
  e^{\frac{i\beta}2 a^T_j \bm\Omega \,a_k},
\end{equation}
one has
\begin{equation}
\bm M_{F*G}^{(\alpha+\beta)}
 =\bm M_F^{(\alpha)}\circ \bm M_G^{(\beta)},
\end{equation}
where $\circ$ denotes the Hadamard product.  The Schur product theorem
implies that this matrix is positive semidefinite.  Thus
$\alpha+\beta\in\N(F*G)$.

\smallskip
\noindent\emph{(4) Convex mixtures.}
If $\eta\in\bigcap_i\N(F_i)$, then
\begin{equation}
\bm M_F^{(\eta)}=\sum_i p_i \bm M_{F_i}^{(\eta)}\geq0.
\end{equation}
For a probability mixture, replace the sum by the corresponding
integral and test against an arbitrary vector $c\in\mathbb C^n$:
$c^* \bm M_F^{(\eta)}c=\int c^* \bm M_{F_s}^{(\eta)}c\,d\mu(s)\geq0$.

\smallskip
\noindent\emph{(5) Dilations.}
A change of variables gives
\begin{equation}
 \widetilde{F_\lambda}(a)=\widetilde F(a \lambda).
\end{equation}
Putting $b_j=\lambda a_j$ and using
$a^T_j \bm \Omega \,a_k   =\lambda^{-2} \, b^T_j \bm \Omega \,b_k$ yields
\begin{equation}
\bm M_{F_\lambda}^{(\eta)}(a_1,\dots,a_n)
 =\bm M_F^{(\eta /  \lambda^2)}(b_1,\dots,b_n).
\end{equation}
The equivalence follows.

\smallskip
\noindent\emph{(6) Tensor products.}
Suppose first that
$\eta\in\N(F_1)\cap\N(F_2)$.  For points
$a_j=(a_j^{(1)},a_j^{(2)})$ in the direct-sum phase space,
\begin{equation}
\bm M_{F_1\otimes F_2}^{(\eta)}
 =\bm M_{F_1}^{(\eta)}(a_1^{(1)},\dots,a_n^{(1)})
  \circ
 \bm M_{F_2}^{(\eta)}(a_1^{(2)},\dots,a_n^{(2)}),
\end{equation}
which is positive semidefinite by the Schur product theorem.  This proves
``$\supseteq$''.  Conversely, restrict the test points to
$(a_j^{(1)},0)$.  Since $\widetilde F_2(0)=1$, the resulting KLM matrix is
exactly $\bm M_{F_1}^{(\eta)}$; hence
$\eta\in\N(F_1)$.  Restriction to $(0,a_j^{(2)})$ gives the same conclusion
for $F_2$.

\smallskip
\noindent\emph{(7) Traces.}
The Fourier transforms of the marginal functions are the restrictions
\begin{equation}
 \widetilde F_1(a_1)=\widetilde F(a_1,0),
 \qquad
 \widetilde F_2(a_2)=\widetilde F(0,a_2).
\end{equation}
If $\eta\in\N(F)$, restricting the KLM test points to $(a_j^{(1)},0)$ or
$(0,a_j^{(2)})$ therefore shows that $\eta$ belongs to the spectrum of
each marginal.

\smallskip
\noindent\emph{(8) Rescaling property II.}
The first equivalence is just a consequence of 5. The characteristic function of a reduced state is obtained
by setting the variables of the traced subsystem equal to zero; hence
$\Theta_\lambda$ commutes with partial traces.  Positivity is preserved by
partial trace, proving the implication.  For a product state,
$\Theta_\lambda[\rho_1\otimes\rho_2]
 =\Theta_\lambda[\rho_1]\otimes\Theta_\lambda[\rho_2]$; because both factors
are Hermitian and have trace one, the tensor product is positive if and
only if both factors are positive.
\end{proof}

\subsection{Narcowich--Wigner spectra of single-mode states}
\label{app:NWsingle}

This section gathers a few tools to compute NW spectra, and the spectra of the single-mode states used in the paper, in particular for numerical results. 

Most of what is known on the subject is due to Br\"ocker and Werner~\cite{broeckerwerner1995}: we recall their results and prove new ones. Recall that for a Hermitian operator $\hat A$ of unit trace, $1\in\N(\hat A)$ if and only if $\hat A\succeq0$, while $0\in\N(\hat A)$ if and only if $W_{\hat A}\ge0$.

\subsubsection{Tools}
\label{sec:NW_tools}

\emph{The rescaling map.} For $\lambda>0$, let $\mathcal L_\lambda$ dilate the Wigner function by $\lambda$,
\begin{align}
    W_{\mathcal L_\lambda[\hat\rho]}(z)=\lambda^{-2}\,W_{\hat\rho}(z/\lambda),
    \qquad
    \chi_{\mathcal L_\lambda[\hat\rho]}(\xi)=\chi_{\hat\rho}(\lambda\xi).
    \label{eq:L_def}
\end{align}
It preserves the trace but not positivity. By the rescaling property, $\N(\mathcal L_\lambda[\hat\rho])=\lambda^2\N(\hat\rho)$, so that
\begin{align}
    \eta\in\N(\hat\rho)
    \quad\Longleftrightarrow\quad
    \mathcal L_{1/\sqrt\eta}[\hat\rho]\succeq0
    \qquad(\eta>0),
    \label{eq:L_positivity}
\end{align}
see Eq.~(3.9) of Ref.~\cite{broeckerwerner1995}; $\mathcal L_{1/\sqrt\eta}[\hat\rho]$ is the operator $\hat\rho_\eta$ of \cref{app:gabor}. Computing a NW spectrum thus amounts to deciding for which $\lambda$ the operator $\mathcal L_\lambda[\hat\rho]$ is positive. The two tools below serve this purpose: a decomposition of $\mathcal L_\lambda$ into Gaussian channels, and its explicit form in the Fock basis.

\emph{Loss and amplification.} Let $\mathcal E_T$ and $\mathcal A_g$ be the pure-loss channel of transmittance $T$ and the quantum-limited amplifier of gain $g$, which act on characteristic functions as
\begin{align}
    \chi_{\mathcal E_T[\hat\rho]}(\xi)=\chi_{\hat\rho}(\sqrt T\,\xi)\,e^{-(1-T)|\xi|^2/4},
    \qquad
    \chi_{\mathcal A_g[\hat\rho]}(\xi)=\chi_{\hat\rho}(\sqrt g\,\xi)\,e^{-(g-1)|\xi|^2/4}.
    \label{eq:loss_amp_chi}
\end{align}
We use these formulas for all $T,g>0$; outside the physical ranges $T\le1$ and $g\ge1$ they define maps that are not positive, but are harmless on states of finite Fock support.

\begin{lemma}[Rescaling as loss followed by amplification]
\label{lem:rescaling_channels}
For every $\lambda>0$,
\begin{align}
    \mathcal L_\lambda=\mathcal A_{(1+\lambda^2)/2}\circ\mathcal E_{2\lambda^2/(1+\lambda^2)} .
    \label{eq:L_decomposition}
\end{align}
\end{lemma}

\begin{proof}
Composing the two formulas of Eq.~\eqref{eq:loss_amp_chi} gives $\chi_{\hat\rho}(\sqrt{gT}\,\xi)\,e^{-(2g-gT-1)|\xi|^2/4}$. The Gaussian factor disappears for $gT=2g-1$, and $gT=\lambda^2$ then fixes $g=(1+\lambda^2)/2$ and $T=2\lambda^2/(1+\lambda^2)$.
\end{proof}

The same computation gives the reverse order, $\mathcal L_\lambda=\mathcal E_{(1+\lambda^2)/2}\circ\mathcal A_{2\lambda^2/(1+\lambda^2)}$. In either form, exactly one of the two channels is unphysical unless $\lambda=1$: rescaling is loss or amplification with the added noise removed. Two consequences follow at once.

First, a positivity test. By Eq.~\eqref{eq:L_positivity}, $\eta\in\N(\hat\rho)$ if and only if $\Tr[\mathcal L_{1/\sqrt\eta}[\hat\rho]\,\hat\sigma]\ge0$ for every pure state $\hat\sigma$. Inserting Eq.~\eqref{eq:L_decomposition} with $\lambda=1/\sqrt\eta$ and moving the amplifier onto $\hat\sigma$ with $\mathcal A_g^\dagger=g^{-1}\mathcal E_{1/g}$, which follows from the Kraus operators of the two channels \cite{ivan2011operator}, we get for $0<\eta\le1$ and $\hat\rho$ of finite Fock support
\begin{align}
    \eta\in\N(\hat\rho)
    \quad\Longleftrightarrow\quad
    \Tr\Big[\mathcal E_{2/(1+\eta)}[\hat\rho]\;\mathcal E_{2\eta/(1+\eta)}[\hat\sigma]\Big]\ge0
    \quad\text{for every pure state }\hat\sigma.
    \label{eq:dual_criterion}
\end{align}

Second, loss fills the spectrum. For $T\in[0,1]$, Eq.~\eqref{eq:L_decomposition} gives $\mathcal L_{1/\sqrt\eta}\circ\mathcal E_T=\mathcal A_{(1+\eta)/2\eta}\circ\mathcal E_{2T/(1+\eta)}$, a composition of two physical channels as soon as $2T-1\le\eta\le1$. For such $\eta$, $\mathcal L_{1/\sqrt\eta}[\mathcal E_T[\hat\rho]]$ is a state, hence
\begin{align}
    \N(\mathcal E_T[\hat\rho])\ \supseteq\ [-1,-(2T-1)]\cup[2T-1,1]
    \qquad\text{for every state }\hat\rho .
    \label{eq:loss_fills}
\end{align}
The two intervals merge at $T=\frac12$: every state is Wigner positive after $50\%$ loss. The point $2T-1$ is also what \cref{thm:single_mode} gives for a beam-splitter of transmittance $T$ fed with $\hat\rho\otimes\ket0\!\bra0$.

\emph{Fock-basis decomposition.} Since $\mathcal L_\lambda$ commutes with phase-space rotations, it preserves the difference $m-n$ of the Fock matrix units,
\begin{align}
    \mathcal L_\lambda[\ket m\!\bra n]=\sum_{r-q=m-n}\Lambda^{(\lambda)}_{mn,rq}\,\ket r\!\bra q .
    \label{eq:L_Fock}
\end{align}
For $m\le n$, integrating the cross-Wigner functions, which are Laguerre polynomials times Gaussians, gives
\begin{align}
    \Lambda^{(\lambda)}_{mn,rq}
    =\frac{2\,(-1)^m}{\lambda^2+1}\sqrt{\frac{n!\,r!}{m!\,q!}}
    \Big(\frac{2\lambda}{\lambda^2+1}\Big)^{n-m}
    \Big(\frac{\lambda^2-1}{\lambda^2+1}\Big)^{m+r}
    \sum_{k=0}^{\min(m,r)}(-1)^k\binom mk\binom q{r-k}\Big(\frac{2\lambda}{1-\lambda^2}\Big)^{2k},
    \label{eq:L_matrix_elements}
\end{align}
and the case $m>n$ follows by Hermitian conjugation. The diagonal coefficients $\Lambda_{nn,qq}$ are Eq.~(3.8) of Ref.~\cite{broeckerwerner1995}, whose parameter $\tau$ is $(\lambda^2-1)/(\lambda^2+1)$; summing them against $z^q$ reproduces the generating function of their Eq.~(3.6), which is how \cref{fig:nw_vs_ncd} is computed. The off-diagonal coefficients are new. Together with Eq.~\eqref{eq:L_positivity}, Eq.~\eqref{eq:L_matrix_elements} reduces the NW spectrum of any state of finite Fock support to the positivity of an explicit matrix, which can be tested numerically by truncation.

\subsubsection{Families of states}
\label{sec:NW_families}

\emph{Passive states.} A single-mode state is passive when it is Fock diagonal with nonincreasing populations, $p_0\ge p_1\ge\dots$ \cite{pusz1978passive,lenard1978thermodynamical}. Such states are Wigner positive \cite{vanherstraeten2021quantum}; in fact more is true.

\begin{lemma}[Passive states]
\label{lem:passive}
Every single-mode passive state satisfies $[-1,1]\subseteq\N(\hat\rho)$.
\end{lemma}

\begin{proof}
Let $\hat\varepsilon_N$ be the uniform mixture of the first $N+1$ Fock states. A passive state is a convex mixture of the $\hat\varepsilon_N$, with weights $(N+1)(p_N-p_{N+1})$ \cite{vanherstraeten2021quantum}, so by convexity and closedness of the spectrum it suffices to treat $\hat\varepsilon_N$. It is the reduced state of the normalised projector $\hat\omega_N$ onto the two-mode subspace of total photon number $N$, which is separable and invariant under every beam-splitter. \Cref{thm:single_mode} applied to $\hat\omega_N$ and a beam-splitter of transmittance $T$ thus gives $2T-1\in\N(\hat\varepsilon_N)$ for every $T\in[0,1]$.
\end{proof}

Conversely, every nonzero operator of finite Fock support has $\N\subseteq[-1,1]$ \cite[Prop.~4]{broeckerwerner1995}, so that a passive state of finite Fock support has exactly $\N(\hat\rho)=[-1,1]$. With an infinite tail the spectrum can be larger: the thermal state of mean photon number $\bar n$ is Gaussian with $\nu=\bar n+\frac12$, so $\N(\hat\rho_{\mathrm{th}})=[-(2\bar n+1),2\bar n+1]$.

\emph{The zero--one sector.} Let
\begin{align}
    \hat\rho=(1-q)\ket0\!\bra0+q\ket1\!\bra1+c\ket0\!\bra1+\bar c\ket1\!\bra0,
    \qquad 0<q\le1,\quad |c|^2\le q(1-q).
    \label{eq:zero_one_state}
\end{align}

\begin{lemma}[Zero--one sector]
\label{lem:zero_one}
The NW spectrum of the state \eqref{eq:zero_one_state} is
\begin{align}
    \N(\hat\rho)=[-1,-\eta_{\min}]\cup[\eta_{\min},1],
    \qquad
    \eta_{\min}=\max\Big\{0,\ \frac{2q^2}{q-|c|^2}-1\Big\}=\max\{0,\ 2\tau_{\hat\rho}-1\}.
    \label{eq:etamin}
\end{align}
\end{lemma}

\begin{proof}
Finite support gives $\N(\hat\rho)\subseteq[-1,1]$, and by symmetry we take $0<\eta\le1$. In the basis $(\ket0,\ket1)$ and for every $T>0$,
\begin{align}
    \mathcal E_T[\hat\rho]=
    \begin{pmatrix}
        1-Tq & \sqrt T\,c\\
        \sqrt T\,\bar c & Tq
    \end{pmatrix}.
    \label{eq:loss_zero_one}
\end{align}
In Eq.~\eqref{eq:dual_criterion} it suffices to take coherent states $\hat\sigma$, which loss maps to coherent states $\ket z$, and $\bra z\hat X\ket z=e^{-|z|^2}(1,\bar z)\hat X(1,z)^T$ for $\hat X$ supported on the sector; the vectors $(1,z)$ exhaust all rays but one, so Eq.~\eqref{eq:dual_criterion} is equivalent to $\mathcal E_{2/(1+\eta)}[\hat\rho]\succeq0$. The determinant of Eq.~\eqref{eq:loss_zero_one} is $T(q-Tq^2-|c|^2)$, so positivity holds if and only if $T\le(q-|c|^2)/q^2$, i.e.\ $\eta\ge2q^2/(q-|c|^2)-1$, which is the first expression of $\eta_{\min}$ (it never exceeds $1$ since $q-|c|^2\ge q^2$).

For the second, note that $\chi^{(-s)}_{\hat\rho}(\xi)=\chi_{\hat\rho}(\xi)e^{-s|\xi|^2/4}$ is, by Eqs.~\eqref{eq:L_def} and \eqref{eq:loss_amp_chi}, the characteristic function of $\mathcal L_{\sqrt{1+s}}\circ\mathcal E_{1/(1+s)}[\hat\rho]$: the smoothed Wigner function $W^{(-s)}_{\hat\rho}$ is nonnegative if and only if $\mathcal E_{1/(1+s)}[\hat\rho]$ is Wigner positive. The Wigner function of Eq.~\eqref{eq:loss_zero_one} is proportional to $1-2Tq+2Tq|z|^2+2\sqrt{2T}\,\mathrm{Re}[c(x+ip)]$, which is nonnegative if and only if $T\le(q-|c|^2)/2q^2$, half the threshold of positivity. Hence $W^{(-s)}_{\hat\rho}\ge0$ if and only if $s\ge2q^2/(q-|c|^2)-1$, i.e.\ $s_{\hat\rho}=-\eta_{\min}$ when $\eta_{\min}>0$, which gives $\eta_{\min}=2\tau_{\hat\rho}-1$; when $\eta_{\min}=0$ the state is Wigner positive and $\tau_{\hat\rho}\le\frac12$.
\end{proof}

In the zero--one sector, the exclusion $|\eta|<2\tau_{\hat\rho}-1$ of \cref{eq:ncdandNW} is thus exactly the gap of the NW spectrum. The diagonal case $c=0$, $\N(\hat\rho)=[-1,-(2q-1)]\cup[2q-1,1]$, is in Sec.~V of Ref.~\cite{broeckerwerner1995}.

\emph{The vacuum--two-photon family.} As soon as $\ket2$ is populated, the NW spectrum carries more information than the nonclassical depth. The family $\hat\rho_p=(1-p)\ket0\!\bra0+p\ket2\!\bra2$ was analysed in Sec.~IV of Ref.~\cite{broeckerwerner1995}, whose Fig.~3 is reproduced and completed in \cref{fig:nw_vs_ncd}: the spectrum is $[-1,1]$ for $p\le\frac13$, develops gaps for $\frac13<p<\frac12$ although the state is still Wigner positive, collapses to the countable set $\{0\}\cup\{\pm(k-\sqrt{k^2-1})\}_{k\ge1}$ at $p=\frac12$ \cite[Eq.~(4.7)]{broeckerwerner1995}, and reduces to $\{\pm1\}$ for $p>\frac12$. The nonclassical depth, $\tau_{\hat\rho_p}=\frac12(1-s_{\hat\rho})$ with $s_{\hat\rho}=\frac{\sqrt{1-p}-\sqrt p}{\sqrt{1-p}+\sqrt p}$, excludes nothing for $p\le\frac12$ and only $|\eta|<2\tau_{\hat\rho_p}-1$ beyond.

\section{Calculations of the examples}

In this section, we provide details for the photon-subtracted examples given in the main text.

\subsection{Probabilistic single-photon subtraction}
\label{app:probabilistic_PSSTS}

The Wigner function of the squeezed thermal state is a positive Gaussian. The photon-subtracted component has the form
\begin{align}
W_{\mathrm{PSSTS}}(x,p)
&\propto
\exp\!\left[
-\frac{e^{2r}x^2+e^{-2r}p^2}{\nu}
\right]
\mathcal{P}(x,p),
\\
\mathcal{P}(x,p)
&=
\frac{(\nu-e^{2r})^2}{\nu}x^2
+
\frac{(\nu-e^{-2r})^2}{\nu}p^2
+
\nu-\cosh(2r).
\label{eq:PSSTS_polynomial}
\end{align}
The quadratic coefficients are non-negative. Hence, the minimum of the polynomial is attained at the origin, where
\begin{equation}
\mathcal{P}(0,0)=\nu-\cosh(2r).
\end{equation}
Consequently, the photon-subtracted state exhibits Wigner negativity if and only if
\begin{equation}
\nu<\cosh(2r).
\label{eq:PSSTS_negativity_condition}
\end{equation}

For the probabilistic mixture, the Wigner function at the origin is a weighted sum of the two component values. Since the squeezed thermal contribution is positive, the negativity threshold is obtained by solving
\begin{equation}
qW_{\mathrm{STS}}(0,0)
+
(1-q)W_{\mathrm{PSSTS}}(0,0)=0.
\end{equation}
Using the normalized component Wigner functions gives
\begin{equation}
q_{\max}
=
\frac{\cosh(2r)-\nu}
{\cosh(2r)(1+\nu^2)-2\nu},
\qquad \nu<\cosh(2r).
\label{eq:PSSTS_qmax_appendix}
\end{equation}
Thus, Wigner negativity survives for $0\leq q<q_{\max}$.

\medskip

\textit{Nonclassical depth.}
The Gaussian-smoothed Wigner function is defined as
\begin{equation}
W_q^{(-s)}(a,b)
=
\frac{1}{\pi s}
\iint_{\mathbb{R}^2}
W_q(x,p)
\exp\!\left[
-\frac{(x-a)^2+(p-b)^2}{s}
\right]
\,dx\,dp.
\label{eq:PSSTS_smoothed_wigner}
\end{equation}
Because the convolution of a Gaussian multiplied by a quadratic polynomial remains a Gaussian multiplied by a quadratic polynomial, the smoothed distribution can be written as
\begin{equation}
W_q^{(-s)}(a,b)
\propto
\exp\!\left[
-\frac{a^2}{s+\nu e^{-2r}}
-\frac{b^2}{s+\nu e^{2r}}
\right]
\left[
C_x(s)a^2+C_p(s)b^2+C_0(s)
\right],
\label{eq:PSSTS_smoothed_form}
\end{equation}
where the coefficients depend on $\nu$, $r$, $q$, and $s$.

The quadratic coefficients are non-negative, and the Gaussian factor is strictly positive. The sign of the smoothed distribution is therefore determined by its value at the origin. The critical smoothing parameter satisfies
\begin{equation}
0=q+\frac{1-q}{\nu\cosh(2r)-1}\Bigg[
1-\frac{\cosh(2r)}{\nu}
+\frac{s\nu e^{-2r}}{2(s+\nu e^{-2r})}
\left(1-\frac{e^{2r}}{\nu}\right)^2
+\frac{s\nu e^{2r}}{2(s+\nu e^{2r})}
\left(1-\frac{e^{-2r}}{\nu}\right)^2
\Bigg].
\label{eq:PSSTS_depth_equation}
\end{equation}
Solving this equation yields the critical smoothing parameter
\begin{equation}
s_{\max}=s_{\max}(\nu,r,q),
\end{equation}
with the physical root chosen within the admissible smoothing interval.
Clearing denominators, Eq.~\eqref{eq:PSSTS_depth_equation} is a quadratic equation in $s$, whose positive root is
\begin{align}
s_{\max}(\nu,r,q)=
\frac{-b_s+\sqrt{b_s^2+4(\nu\cosh 2r-1)\big[(1-q)\nu(\cosh 2r-\nu)-q\nu^2(\nu\cosh 2r-1)\big]}}{2(\nu\cosh 2r-1)},
\label{eq:PSSTS_depth_explicit}
\\[1em]
b_s=2q\nu\cosh 2r\,(\nu\cosh 2r-1)+(1-q)(\nu^2-1).
\nonumber
\end{align}
The constant term of the quadratic is negative exactly when $q<q_{\max}$; for $q\geq q_{\max}$ the Wigner function is nonnegative and we set $s_{\max}=0$. The nonclassical depth is $\tau_{\hat\rho_q}=(1+s_{\max})/2$, and for $\nu=1$ the root reduces to $s_{\max}=\sqrt{q^2\cosh^2 2r+1-2q}-q\cosh 2r$.

\medskip

\textit{Characteristic function and $\eta$-positivity.}
In the variables $\tilde x=e^{r}\xi_x$ and $\tilde p=e^{-r}\xi_p$, which undo the squeezing, the characteristic function rescaled by $1/\sqrt{|\eta|}$, which by Eq.~\eqref{eq:L_positivity} must be that of a positive operator for $\eta$ to belong to the NW spectrum, reads
\begin{equation}
\chi_q(\tilde{x},\tilde{p})
\propto
\exp\!\left[
-\frac{\nu}{4|\eta|}(\tilde{x}^2+\tilde{p}^2)
\right]
\left(
1-A_q\tilde{x}^2-B_q\tilde{p}^2
\right).
\label{eq:PSSTS_chi_appendix}
\end{equation}
The coefficients $A_q$ and $B_q$ incorporate the relative weight of the photon-subtracted contribution and the normalization of the probabilistic preparation. They read
\begin{align}
A_q
&=
\frac{1}{|\eta|}
(1-q)
\frac{e^{-2r}(\nu e^{2r}-1)^2}
{4[\nu\cosh(2r)-1]},
\label{eq:PSSTS_Aq}\\
B_q
&=
\frac{1}{|\eta|}
(1-q)
\frac{e^{2r}(\nu e^{-2r}-1)^2}
{4[\nu\cosh(2r)-1]}.
\label{eq:PSSTS_Bq}
\end{align}

The $\eta$-positivity condition requires the matrix
\begin{equation}
\left[
e^{\frac{i}{2}(x_jp_k-x_kp_j)}
\chi_q(x_k-x_j,\,p_k-p_j)
\right]_{j,k}
\succeq 0
\label{eq:PSSTS_eta_positivity}
\end{equation}
for every finite collection of phase-space points. This is the requirement that the rescaled characteristic function be of $1$-positive type, or equivalently, the characteristic function of a positive operator.

Following the operator construction, introduce
\begin{equation}
\hat{\mathcal{K}}_q
=
\hat{\rho}_{\mathrm{th}}
-\frac{A_q-B_q}{\mu^2+1}
\left\{
\hat{a}^2+\hat{a}^{\dagger 2},
\hat{\rho}_{\mathrm{th}}
\right\}
+2(A_q+B_q)\partial_\mu\hat{\rho}_{\mathrm{th}},
\label{eq:PSSTS_Kprime}
\end{equation}
where $\hat\rho_{\mathrm{th}}$ is the thermal state of parameter $\mu=\nu/|\eta|$, i.e.\ $\chi_{\hat\rho_{\mathrm{th}}}(\xi)=e^{-\mu|\xi|^2/4}$, and $\partial_\mu$ is the derivative with respect to this parameter. Since $\chi_{\{\hat a^2+\hat a^{\dagger 2},\hat\rho_{\mathrm{th}}\}}(\xi)=\tfrac{\mu^2+1}{2}(\xi_x^2-\xi_p^2)\,\chi_{\hat\rho_{\mathrm{th}}}(\xi)$ and $\chi_{\partial_\mu\hat\rho_{\mathrm{th}}}(\xi)=-\tfrac{|\xi|^2}{4}\chi_{\hat\rho_{\mathrm{th}}}(\xi)$, its characteristic function is exactly the right-hand side of Eq.~\eqref{eq:PSSTS_chi_appendix}.

We factorize the corresponding operator as
\begin{equation}
\hat{\mathcal{K}}_q
=
\hat{\rho}_{\mathrm{th}}^{1/2}
\hat{\Sigma}_q
\hat{\rho}_{\mathrm{th}}^{1/2},
\end{equation}
where
\begin{equation}
\hat{\Sigma}_q
=
\alpha_q+\beta_q\hat{n}
-\gamma_q(\hat{a}^2+\hat{a}^{\dagger 2}).
\label{eq:PSSTS_Sigma}
\end{equation}
The coefficients are determined by matching the characteristic functions. Since the thermal state is full rank for $\mu>1$, i.e.\ $|\eta|<\nu$, positivity of $\hat{\mathcal{K}}_q$ is equivalent to positivity of $\hat{\Sigma}_q$.

To diagonalize the quadratic part, define the canonical quadratures
\begin{equation}
\hat{X}
=
g\frac{\hat{a}+\hat{a}^{\dagger}}{\sqrt{2}},
\qquad
\hat{P}
=
\frac{1}{g}
\frac{\hat{a}-\hat{a}^{\dagger}}{i\sqrt{2}},
\qquad
[\hat{X},\hat{P}]=i.
\end{equation}
Choosing the rescaling to equalize the coefficients of $\hat{X}^2$ and $\hat{P}^2$ reduces the operator to
\begin{equation}
\hat{\Sigma}_q
=
V_0(q)(\hat{X}^2+\hat{P}^2)+c_0(q),
\end{equation}
where
\begin{align}
g^2&=\sqrt{\frac{B_q}{A_q}},\\
V_0(q)&=\frac{4\sqrt{A_qB_q}}{\mu^2-1},\\
c_0(q)&=
1-\frac{2(A_q+B_q)}{\mu+1}
-\frac{2(A_q+B_q)}{\mu^2-1}.
\end{align}
The lowest eigenvalue of $\hat{X}^2+\hat{P}^2$ is $1$. Hence, positivity requires
\begin{equation}
V_0(q)+c_0(q)\geq0,
\end{equation}
or, equivalently,
\begin{equation}
2\mu(A_q+B_q)-4\sqrt{A_qB_q}
\leq \mu^2-1.
\label{eq:PSSTS_positivity_inequality}
\end{equation}

Substituting Eqs.~\eqref{eq:PSSTS_Aq} and
\eqref{eq:PSSTS_Bq} into Eq.~\eqref{eq:PSSTS_positivity_inequality}
and solving the resulting quadratic inequality in $|\eta|$ yields
\begin{equation}
|\eta|\in
[\eta_-(\nu,r,q),\eta_+(\nu,r,q)].
\label{eq:PSSTS_eta_interval_appendix}
\end{equation}
The endpoints are the two roots of the equality in Eq.~\eqref{eq:PSSTS_positivity_inequality}, which with $\mu=\nu/|\eta|$ and $A_q=a_q/|\eta|$, $B_q=b_q/|\eta|$ is a quadratic equation in $|\eta|$:
\begin{equation}
\eta_\pm(\nu,r,q)=2\sqrt{a_qb_q}\pm\sqrt{(\nu-2a_q)(\nu-2b_q)} .
\label{eq:PSSTS_eta_pm}
\end{equation}
The radicand is nonnegative since $2a_q,2b_q\leq\nu$, and $\eta_+\leq\nu$ by the Cauchy--Schwarz inequality, with equality for $q=1$, where the state is Gaussian. The derivation holds for $|\eta|<\nu$, and we have checked numerically, with the Fock-basis form of the rescaling map of \cref{sec:NW_tools}, that no larger value belongs to the spectrum. One has $\eta_-\leq0$ exactly when $q\geq q_{\max}$, consistently with Bochner's theorem, and for $q=0$ the radicand equals $(\nu^2-1)^2/4(\nu\cosh 2r-1)^2$, which gives $\eta_+=1$ and $\eta_-=\nu(\cosh 2r-\nu)/(\nu\cosh 2r-1)$, the positive root of Eq.~\eqref{eq:PSSTS_depth_equation} at $q=0$.

\subsection{Two-photon-subtracted single-mode squeezed vacuum state with probabilistic success}
\label{app:probabilistic_2PSSTS}
We numerically determine the NW spectrum of the probabilistic two-photon-subtracted squeezed vacuum by exploiting the rescaling map introduced in \cref{app:NWsingle}. In particular, the rescaling property allows us to reformulate the $\eta$-positivity condition as an operator-positivity problem:
\begin{equation}
    \eta\in \N(F_{\hat\rho_{\mathrm{2PSSV}}})
    \quad\Longleftrightarrow\quad
    \mathcal L_{1/\sqrt{\eta}}[\hat\rho_{\mathrm{2PSSV}}]\succeq0,
    \qquad \eta>0.
\end{equation}
We therefore construct the matrix elements of the rescaled operator in the Fock basis using the explicit expression of the rescaling-map coefficients. Since the input state is supported on the vacuum and two-photon Fock states, its density operator contains only the four contributions
$\{|0\rangle\langle0|,|2\rangle\langle2|,|0\rangle\langle2|,|2\rangle\langle0|\}$.
Under the action of $\mathcal L_\lambda$, these contributions generate operators in the span of
\begin{equation}
    \left\{
    |n\rangle\langle n|,
    |n+2\rangle\langle n+2|,
    |n\rangle\langle n+2|,
    |n+2\rangle\langle n|
    \right\}_{n=0}^{\infty}.
\end{equation}
In other words, $\mathcal L_\lambda$ adds or removes the same number of photons on the ket and bra sides, thereby preserving the photon-number difference of each Fock-basis matrix element. Although the input state has finite Fock support, its rescaled image generally occupies infinitely many Fock levels. We thus consider finite-dimensional principal truncations of the resulting matrix and evaluate their smallest eigenvalues as functions of $\lambda$. Since the rescaled matrix contains only diagonal and second off-diagonal elements, it decomposes into independent even- and odd-photon-number blocks.

The positive values of $\lambda$ for which both blocks are positive semidefinite determine the positive part of the NW spectrum through $\eta=\lambda^{-2}$. The negative part follows from the symmetry $\N(F_{\hat\rho})=-\N(F_{\hat\rho})$ for Hermitian operators.

Positivity of a finite submatrix does not guarantee positivity of the full infinite-dimensional operator, although a negative eigenvalue certifies its non-positivity. We therefore assess numerical convergence by increasing the Fock-space cutoff and monitoring the stability of the positivity boundaries. 

\section{Proofs of \cref{thm:single_mode,thm:gaussian}}
\label{app:proofmain}

In this section we prove the following multimode generalization of \cref{thm:single_mode}, and also give a proof of \cref{thm:gaussian}. 

\medskip

The generalization involves a positivity property of characteristic functions, the $\bm M$-positive type of \cref{def:Mpositive} below, which reduces to the NW spectrum when a single output mode is retained.

\begin{theorem}[Multimode version of \cref{thm:single_mode}]
\label{thm:main}
Let $\hat\rho$ be an $N$-mode state and $A|B$ a bipartition of its modes. 
Let $\hat U_G$ be a Gaussian unitary with symplectic matrix $\bm S$, let $O$ be a subset of the output modes, and let $\hat\sigma=\Tr_{\bar O}[\hat U_G\hat\rho\hat U_G^\dagger]$ be the reduced output state on $O$. 
Denote by $\bm S_{OA}$ and $\bm S_{OB}$ the blocks of $\bm S$ mapping the quadratures of $A$ and of $B$ onto those of $O$, let $\bm K_A=\bm S_{OA}\bm\Omega_A\bm S_{OA}^T$ and $\bm K_B=\bm S_{OB}\bm\Omega_B\bm S_{OB}^T$, so that $\bm K_A+\bm K_B=\bm\Omega_O$, the symplectic matrix of the modes in $O$. 
If $\hat\rho$ is PPT across $A|B$, then the characteristic function $\chi_{\hat\sigma}$ is of $\bm\Omega_O^T(\bm K_A-\bm K_B)\bm\Omega_O$-positive type.
\end{theorem}

The idea of the proof is simple. 
Positivity of an operator constrains its characteristic function through the Kastler--Loupias--Miracle-Sol\'e (KLM) conditions, a twisted version of Bochner's theorem whose phase is set by the commutators of the observables involved. The characteristic function of the output modes can be read either as that of $\hat\rho$ with respect to the output quadratures, or as that of $\hat\rho^{\Gamma_B}$ with respect to ``mirror'' quadratures with different commutators; for a PPT state both operators are positive, and $\chi_{\hat\sigma}$ inherits both KLM conditions. 
We recall the KLM conditions for arbitrary commutation relations in \cref{sec:KLM}, establish the two readings of $\chi_{\hat\sigma}$ in \cref{sec:gaussian_setup,sec:PT_reversal}, and conclude in \cref{sec:proof_main,sec:single_mode_case}.

\subsection{Positivity and the KLM conditions}
\label{sec:KLM}

Let $\hat R=(\hat R_1,\dots,\hat R_{2m})^T$ be Hermitian operators whose commutators are $c$-numbers,
\begin{align}
    [\hat R,\hat R^T]=i\bm K ,
    \label{eq:Kcommutator}
\end{align}
where $\bm K$ is a real antisymmetric $2m\times2m$ matrix. We do not assume that $\bm K$ equals the canonical matrix $\bm\Omega$, nor that it is invertible. In this appendix, $\hat R$ will always be a real linear combination $\hat R=\bm L\hat r$ of the canonical quadratures of some set of modes, in which case $\bm K=\bm L\bm\Omega\bm L^T$. If $\bm L$ collects some rows of a symplectic matrix, $\bm K$ is the corresponding block of $\bm\Omega$; other choices of $\bm L$ produce other, possibly degenerate, antisymmetric matrices.

The Weyl operators associated with $\hat R$ are $\hat D_{\bm K}(\xi)=e^{i\hat R^T\bm\Omega_m\xi}$, $\xi\in\reals^{2m}$, where $\bm{\Omega}_{m}$ is the canonical symplectic matrix of size $2m$.
They satisfy the Weyl relation
\begin{align}
    \hat D_{\bm K}(\xi)\,\hat D_{\bm K}(\zeta)^\dagger
    =e^{\frac i2\,\xi^T\bm\Omega_m^T\bm K\bm\Omega_m\,\zeta}\,\hat D_{\bm K}(\xi-\zeta),
    \label{eq:weyl}
\end{align}
which follows from the Baker--Campbell--Hausdorff formula, since the commutator of $i\hat R^T\bm\Omega_m\xi$ and $-i\hat R^T\bm\Omega_m\zeta$ is $i(\bm\Omega_m\xi)^T\bm K(\bm\Omega_m\zeta)$. 
For $\hat R=\bm L\hat r$ one has $\hat D_{\bm K}(\xi)=e^{i\hat r^T\bm L^T\bm\Omega_m\xi}=\hat D(\bm\Omega^T\bm L^T\bm\Omega_m\xi)$, an ordinary Weyl operator of the underlying modes; for $\bm K=\bm\Omega_m$, Eq.~\eqref{eq:weyl} is the usual Weyl relation, since $\bm\Omega_m^T\bm\Omega_m\bm\Omega_m=\bm\Omega_m$.

For a trace-class operator $\hat X$, we define its characteristic function with respect to $\hat R$ as
\begin{align}
    \chi^{\hat R}_{\hat X}(\xi)=\Tr\big[\hat X\,\hat D_{\bm K}(\xi)\big],\qquad\xi\in\reals^{2m}.
    \label{eq:chi_R}
\end{align}
It is a bounded continuous function with $\chi^{\hat R}_{\hat X}(0)=\Tr\hat X$, and it satisfies $\chi^{\hat R}_{\hat X}(-\xi)=\overline{\chi^{\hat R}_{\hat X}(\xi)}$ when $\hat X$ is Hermitian. 
When $\hat R=\hat r$ and $\hat X=\hat\rho$ is a state, it is the usual characteristic function $\chi_{\hat\rho}$ of Eq.~\eqref{eq:chi_def}.

The positivity condition that we need is the following.

\begin{definition}[$\bm M$-positive type]
\label{def:Mpositive}
Let $\bm M$ be a real antisymmetric $2m\times2m$ matrix. A function $\chi:\reals^{2m}\to\complex$ is of \emph{$\bm M$-positive type} if, for every finite family of points $\xi_1,\dots,\xi_n\in\reals^{2m}$, the $n\times n$ matrix
\begin{align}
    \Big[\chi(\xi_a-\xi_b)\,e^{\frac i2\,\xi_a^T\bm M\xi_b}\Big]_{a,b=1}^{n}
    \label{eq:Mpositive}
\end{align}
is positive semidefinite.
\end{definition}

For $\bm M=0$ the phase disappears and \cref{def:Mpositive} is the usual notion of a positive-definite function, which by Bochner's theorem characterises the Fourier transforms of positive measures. For $\bm M=\bm\Omega$ it is the twisted positive-definiteness appearing in the KLM theorem~\cite{Kastler1965,Loupias1966,Loupias1967}; for $\bm M=\eta\bm\Omega$ it is Narcowich's $\eta$-positive type~\cite{narcowich_conditions_1988,narcowich_distributions_1989}; for a general antisymmetric $\bm M$ it is the $(1,\bm M)$-positive type of Dias and Prata~\cite{diasprata2018}. 

The link between the positivity of operators and the $\bm M$-positivity of functions is the following elementary statement, which is the only part of the KLM theory that we need.

\begin{lemma}[Positivity implies twisted positivity]
\label{lem:KLM}
Let $\hat R$ satisfy Eq.~\eqref{eq:Kcommutator}, and let $\hat X$ be a positive semidefinite trace-class operator. 
Then $\chi^{\hat R}_{\hat X}$ is of $\bm\Omega_m^T\bm K\bm\Omega_m$-positive type.
\end{lemma}
\begin{proof}
Since $\hat X\succeq0$, we have $\Tr[\hat X\hat Y\hat Y^\dagger]\ge0$ for every bounded operator $\hat Y$. Take $\hat Y=\sum_a\bar c_a\hat D_{\bm K}(\xi_a)$, for arbitrary points $\xi_1,\dots,\xi_n\in\reals^{2m}$ and coefficients $c_1,\dots,c_n\in\complex$. Expanding $\hat Y\hat Y^\dagger$ and using the Weyl relation~\eqref{eq:weyl} to reorder each product $\hat D_{\bm K}(\xi_a)\hat D_{\bm K}(\xi_b)^\dagger$, we obtain
\begin{align}
    0\le\Tr[\hat X\hat Y\hat Y^\dagger]
    =\sum_{a,b}\bar c_a c_b\,e^{\frac i2\,\xi_a^T\bm\Omega_m^T\bm K\bm\Omega_m\xi_b}\,\chi^{\hat R}_{\hat X}(\xi_a-\xi_b),
    \label{eq:KLM}
\end{align}
which is the positive semidefiniteness of the matrix~\eqref{eq:Mpositive} with $\bm M=\bm\Omega_m^T\bm K\bm\Omega_m$.
\end{proof}

Thus the phase in \cref{def:Mpositive} is the one produced when the Weyl operators are reordered, and it is entirely determined by the commutator matrix $\bm K$. 
In particular, the same operator $\hat X$, looked at through two families of observables with different commutators, yields characteristic functions of different positive types. 
This is the mechanism at work in the proof of \cref{thm:main}.

Note that the sign of $\bm M$ is irrelevant for Hermitian functions: if $\chi(-\xi)=\overline{\chi(\xi)}$ for all $\xi$, then
\begin{align}
    \chi\ \text{is of }\bm M\text{-positive type}
    \quad\Longleftrightarrow\quad
    \chi\ \text{is of }(-\bm M)\text{-positive type}.
    \label{eq:sign_symmetry}
\end{align}
Indeed, evaluating the matrix~\eqref{eq:Mpositive} at the points $-\xi_1,\dots,-\xi_n$ and using $\chi(-\xi)=\overline{\chi(\xi)}$ yields the complex conjugate of the matrix associated with $-\bm M$ and the points $\xi_1,\dots,\xi_n$, and the complex conjugate of a positive semidefinite matrix is positive semidefinite. 
In particular, the ordering of the indices $a,b$ in the phase of Eq.~\eqref{eq:Mpositive}, which differs among the references cited above, is a matter of convention. 
Equation~\eqref{eq:sign_symmetry} will also account for the symmetric roles of $A$ and $B$ in \cref{thm:main}.

Finally, let us connect \cref{def:Mpositive} with the NW spectrum. 
For a single mode, $m=1$, every real antisymmetric matrix is a multiple of $\bm J$, 
and the NW spectrum of a state $\hat\sigma$ is precisely the set of values of $\eta$ for which its characteristic function satisfies the twisted positivity condition of \cref{def:Mpositive} [\cref{sec:background_NW}, Refs.~\cite{narcowich_conditions_1988,broeckerwerner1995}]:
\begin{align}
    \eta\in\N(\hat\sigma)
    \quad\Longleftrightarrow\quad
    \chi_{\hat\sigma}\ \text{is of }\eta\bm J\text{-positive type}.
    \label{eq:NW_positive_type}
\end{align}
For $\eta>0$, by the KLM theorem~\cite{Kastler1965,Loupias1966,Narcowich1986}
 with Planck constant $\eta$, this means that $W_{\hat\sigma}$ is a bona fide Wigner function when $\hbar=\eta$, i.e.\ that the operator $\hat\sigma_\eta$ of \cref{app:gabor} is positive semidefinite; for $\eta=0$ it is Bochner's condition, $W_{\hat\sigma}\ge0$. 

In this language, \cref{lem:KLM} with $\bm K=\bm J$ (for which $\bm J^T\bm J\bm J=\bm J$) is the trivial statement that $1\in\N(\hat\sigma)$ for every state, and Eq.~\eqref{eq:sign_symmetry} is the symmetry $\N(\hat\sigma)=-\N(\hat\sigma)$. 
 When several output modes are retained, the commutator matrices that appear are no longer multiples of $\bm\Omega_O$, and $\bm M$-positivity is the appropriate language; this is why \cref{thm:main} is stated in these terms.

\subsection{Gaussian processing and the two contributions to the output commutator}
\label{sec:gaussian_setup}

We now turn to the setting of \cref{thm:main}. An $N$-mode state $\hat\rho$, with a bipartition $A|B$ of its modes, is processed by a Gaussian unitary $\hat U_G$ with symplectic matrix $\bm S$; a set $O$ of output modes is retained and its complement $\bar O$ is discarded,
\begin{align}
    \hat\sigma=\Tr_{\bar O}\big[\hat U_G\,\hat\rho\,\hat U_G^\dagger\big].
    \label{eq:sigma_def}
\end{align}
In the Heisenberg picture, the output quadratures are $\hat q\coloneqq\hat U_G^\dagger\hat r\hat U_G=\bm S\hat r$; being related to the input quadratures by a symplectic transformation, they obey the same commutation relations, $[\hat q,\hat q^T]=i\bm S\bm\Omega\bm S^T=i\bm\Omega$. 
The characteristic function of $\hat\sigma$ is obtained by evaluating the Weyl operators of the retained modes on the output state; since these operators act only on the modes in $O$, the partial trace over $\bar O$ plays no role, and passing to the Heisenberg picture gives
\begin{align}
    \chi_{\hat\sigma}(x)
    =\Tr\big[\hat\sigma\,e^{i\hat r_O^T\bm\Omega_O x}\big]
    =\Tr\big[\hat\rho\,e^{i\hat q_O^T\bm\Omega_O x}\big]
    =\chi^{\hat q_O}_{\hat\rho}(x),
    \qquad x\in\reals^{2|O|},
    \label{eq:chi_sigma}
\end{align}
where $\hat q_O$ collects the quadratures of the retained modes.

Let us order the input modes as $(A,B)$ and the output modes as $(O,\bar O)$, and write
\begin{align}
    \bm S=\begin{pmatrix}\bm S_{OA}&\bm S_{OB}\\ \bm S_{\bar OA}&\bm S_{\bar OB}\end{pmatrix},
    \qquad
    \hat q_O=\bm S_{OA}\hat r_A+\bm S_{OB}\hat r_B .
    \label{eq:qO_split}
\end{align}
Each retained quadrature is thus the sum of a contribution from $A$ and a contribution from $B$. 
Since the quadratures of $A$ commute with those of $B$, and since $\bm\Omega=\bm\Omega_A\oplus\bm\Omega_B$ in the ordering $(A,B)$, the commutator matrix of $\hat q_O$ splits accordingly,
\begin{align}
    [\hat q_O,\hat q_O^T]=i\,(\bm K_A+\bm K_B),
    \qquad
    \bm K_A\coloneqq\bm S_{OA}\bm\Omega_A\bm S_{OA}^T,
    \qquad
    \bm K_B\coloneqq\bm S_{OB}\bm\Omega_B\bm S_{OB}^T .
    \label{eq:KAKB}
\end{align}
Here $\bm K_A$ is the commutator matrix of the operators $\bm S_{OA}\hat r_A$ alone, i.e.\ of the part of the output quadratures that comes from $A$, and similarly for $\bm K_B$. 
On the other hand, $[\hat q_O,\hat q_O^T]$ is the $(O,O)$ block of $i\bm\Omega$, so that
\begin{align}
    \bm K_A+\bm K_B=\bm\Omega_O .
    \label{eq:split}
\end{align}
This sum rule is simply the symplecticity of $\bm S$ restricted to the retained modes: the two contributions add up to the canonical commutator of the output quadratures. 
\Cref{thm:main} involves their difference.

It is useful to see what these matrices become when a single output mode is retained, which is the situation of \cref{thm:single_mode}. 
Let $O=\{b\}$ and let $\hat b=\sum_k(u_k\hat a_k+v_k\hat a_k^\dagger)$ be the corresponding output mode, as in Eq.~\eqref{eq:b}. 
The block $\bm S_{OA}$ is now a row of $2\times2$ blocks $\bm S_{Ok}$, $k\in A$, where $\bm S_{Ok}$ maps the quadratures of input mode $k$ onto those of $\hat b$, and
\begin{align}
    \bm K_A=\sum_{k\in A}\bm S_{Ok}\bm J\bm S_{Ok}^T
    =\Big(\sum_{k\in A}\det\bm S_{Ok}\Big)\bm J ,
    \label{eq:KA_single}
\end{align}
where we used the identity $\bm M\bm J\bm M^T=\det(\bm M)\,\bm J$, valid for any real $2\times2$ matrix $\bm M$. The determinant $\det\bm S_{Ok}$ has a direct meaning. 
The contribution of input mode $k$ to $\hat b$ is $\hat c_k=u_k\hat a_k+v_k\hat a_k^\dagger$, whose quadratures are $\bm S_{Ok}\hat r_k$; computing the commutator of these two quadratures either from $\bm S_{Ok}\bm J\bm S_{Ok}^T$ or from $\hat c_k$ gives $\det\bm S_{Ok}=[\hat c_k,\hat c_k^\dagger]=|u_k|^2-|v_k|^2=w_k$, the weight of mode $k$ in the output mode. 
Hence
\begin{align}
    \bm K_A=w_A\bm J,\qquad\bm K_B=w_B\bm J,\qquad w_A\coloneqq\sum_{k\in A}w_k,\quad w_B\coloneqq\sum_{k\in B}w_k,
    \label{eq:weights}
\end{align}
and the sum rule~\eqref{eq:split} reads $w_A+w_B=1$, which is the normalisation $[\hat b,\hat b^\dagger]=1$ of Eq.~\eqref{eq:b}. 
For a beam-splitter of transmittance $\eta$ mixing one mode of $A$ with one mode of $B$, $w_A=\eta$; for a two-mode squeezer of gain $g=\cosh^2r$, $w_A=g$ and $w_B=1-g<0$: active transformations can make one of the two contributions negative.

\subsection{Partial transposition reverses the contribution of $B$}
\label{sec:PT_reversal}

We now bring in the partial transposition. 
Recall from \cref{sec:background_PPT} that transposing $B$ acts on the quadratures as the reflection $\hat{\mathbf r}\mapsto\bm\Lambda_B\hat{\mathbf r}$ of the momenta of $B$ (see \cref{eq:LambdaB}). Since Weyl operators factorise over $A$ and $B$ and $(e^{i\hat A})^T=e^{i\hat A^T}$ for Hermitian $\hat A$, partial transposition acts on the Weyl operators of the input modes as
\begin{align}
    \big(e^{i\hat{\mathbf r}^Tu}\big)^{\Gamma_B}=e^{i(\bm\Lambda_B\hat{\mathbf r})^Tu}=e^{i\hat{\mathbf r}^T\bm\Lambda_Bu},
    \qquad u\in\reals^{2N},
    \label{eq:PT_weyl}
\end{align}
i.e.\ $\hat D(z)^{\Gamma_B}=\hat D(\bm\Omega^T\bm\Lambda_B\bm\Omega z)$: in terms of displacements, it reflects the positions of $B$. 
Moreover $\Tr[\hat X\hat Y^{\Gamma_B}]=\Tr[\hat X^{\Gamma_B}\hat Y]$ for any two operators.

Consider the quadratures
\begin{align}
    \hat q'\coloneqq\bm S\bm\Lambda_B\hat r ,
    \qquad
    \hat q'_O=\bm S_{OA}\hat r_A+\bm S_{OB}\bm\Lambda_B\hat r_B ,
    \label{eq:qprime}
\end{align}
that is, the output quadratures that the Gaussian unitary would produce if it were fed with the reflected input quadratures $\bm\Lambda_B\hat r$: the contribution of $A$ is unchanged, and in the contribution of $B$ every momentum is replaced by its opposite (in the second expression, $\bm\Lambda_B$ denotes the block of the reflection acting on the modes of $B$). 
These are Hermitian linear combinations of canonical quadratures, so they fall within the scope of \cref{sec:KLM}, but their commutation relations are not canonical. 
Indeed, reflecting the momenta of $B$ reverses the sign of the commutators within $B$, $\bm\Lambda_B\bm\Omega\bm\Lambda_B=\bm\Omega_A\oplus(-\bm\Omega_B)$, and the same computation as in Eq.~\eqref{eq:KAKB} gives
\begin{align}
    [\hat q'_O,\hat q'^{T}_O]=i\,(\bm K_A-\bm K_B).
    \label{eq:reversed_share}
\end{align}
Reversing the momenta of $B$ before the Gaussian processing leaves the contribution of $A$ to the output commutator untouched and reverses the contribution of $B$. 
Note that $\bm K_A-\bm K_B$ may well be degenerate, as happens for a balanced beam-splitter, where $\bm K_A=\bm K_B=\bm J/2$; this is why \cref{lem:KLM} was stated for an arbitrary antisymmetric $\bm K$.

The characteristic function of the output can now be read in two ways. 
The operator $e^{i\hat q_O^T\bm\Omega_Ox}=e^{i\hat r^Tu}$ with $u=\bm S_O^T\bm\Omega_Ox$, where $\bm S_O=(\bm S_{OA}\;\;\bm S_{OB})$ denotes the $O$ rows of $\bm S$, is a Weyl operator of the $N$ input modes. 
By Eq.~\eqref{eq:PT_weyl}, its partial transpose is $e^{i\hat r^T\bm\Lambda_Bu}=e^{i(\bm S_O\bm\Lambda_B\hat r)^T\bm\Omega_Ox}=e^{i\hat q'^T_O\bm\Omega_Ox}$, the Weyl operator of the mirror quadratures $\hat q'_O$ labelled by the same displacement $x$.
Moving the partial transposition from the Weyl operator to the state in Eq.~\eqref{eq:chi_sigma}, we obtain
\begin{align}
    \chi_{\hat\sigma}(x)
    =\Tr\big[\hat\rho\,e^{i\hat q_O^T\bm\Omega_Ox}\big]
    =\Tr\big[\hat\rho^{\Gamma_B}\,e^{i\hat q'^T_O\bm\Omega_Ox}\big]
    =\chi^{\hat q'_O}_{\hat\rho^{\Gamma_B}}(x).
    \label{eq:bridge}
\end{align}
Equation~\eqref{eq:bridge} is the heart of the proof. 
Its left-hand side is the characteristic function of the retained output modes, a measurable quantity. 
The first expression reads it as the characteristic function of $\hat\rho$ with respect to the output quadratures $\hat q_O$, whose commutator matrix is $\bm K_A+\bm K_B=\bm\Omega_O$; the second reads it as the characteristic function of the operator $\hat\rho^{\Gamma_B}$ with respect to the mirror quadratures $\hat q'_O$, whose commutator matrix is $\bm K_A-\bm K_B$. 
Nothing has been assumed about $\hat\rho$ so far: Eq.~\eqref{eq:bridge} holds for every state, entangled or not, and $\hat\rho^{\Gamma_B}$ is always a Hermitian operator. 
What distinguishes PPT states is that $\hat\rho^{\Gamma_B}$ is, in addition, positive.

\subsection{Proof of \cref{thm:main}}
\label{sec:proof_main}

\begin{proof}[Proof of \cref{thm:main}]
Suppose that $\hat\rho$ is PPT across $A|B$, i.e.\ $\hat\rho^{\Gamma_B}\succeq0$. 
Then $\hat\rho^{\Gamma_B}$ is a density operator. 
We may therefore apply \cref{lem:KLM} to the second reading of Eq.~\eqref{eq:bridge}, with $\hat X=\hat\rho^{\Gamma_B}$, $\hat R=\hat q'_O$ and, by Eq.~\eqref{eq:reversed_share}, $\bm K=\bm K_A-\bm K_B$: the function $\chi_{\hat\sigma}=\chi^{\hat q'_O}_{\hat\rho^{\Gamma_B}}$ is of $\bm\Omega_O^T(\bm K_A-\bm K_B)\bm\Omega_O$-positive type, which is the statement of \cref{thm:main}. 
By the sum rule~\eqref{eq:split} and $\bm\Omega_O^T\bm\Omega_O\bm\Omega_O=\bm\Omega_O$, this matrix also reads $2\bm\Omega_O^T\bm K_A\bm\Omega_O-\bm\Omega_O$.
\end{proof}

A few comments are in order. 
First, applying \cref{lem:KLM} to the first reading of Eq.~\eqref{eq:bridge} instead, with $\hat X=\hat\rho$ and $\bm K=\bm\Omega_O$, only tells us that $\chi_{\hat\sigma}$ is of $\bm\Omega_O$-positive type, which merely expresses that $\hat\sigma$ is a quantum state and holds for every input: all the information lies in the second reading. 
Second, \cref{thm:main} treats $A$ and $B$ asymmetrically only in appearance. 
By \cref{eq:sign_symmetry}, since $\chi_{\hat\sigma}$ is the characteristic function of a Hermitian operator, $\bm{\Omega}_{O}^{T}(\bm K_A-\bm K_B)\bm{\Omega}_{O}$-positivity is equivalent to $\bm{\Omega}_{O}^{T}(\bm K_B-\bm K_A)\bm{\Omega}_{O}$-positivity, which is what one obtains by transposing $A$ instead of $B$. 
This is as it should be, since $\hat\rho^{\Gamma_A}=(\hat\rho^{\Gamma_B})^T$ and the PPT property does not depend on which party is transposed. 
Third, \cref{thm:main} provides, for every choice of Gaussian unitary and of retained modes, a necessary condition for $\hat\rho$ to be PPT, hence a sufficient condition for NPT entanglement, which involves only the reduced output state $\hat\sigma$.

\subsection{A single output mode: back to the Narcowich--Wigner spectrum}
\label{sec:single_mode_case}

We finally retain a single output mode $\hat b$, as in \cref{thm:single_mode}. 
By Eq.~\eqref{eq:weights}, $\bm K_A-\bm K_B=(w_A-w_B)\bm J=(2w_A-1)\bm J$ and $\bm\Omega_O=\bm J$, so that
\begin{align}
    \bm\Omega_O^T(\bm K_A-\bm K_B)\bm\Omega_O=(2w_A-1)\,\bm J^T\bm J\bm J=(2w_A-1)\,\bm J :
\end{align}
for a single mode \cref{thm:main} states that, if $\hat\rho$ is PPT across $A|B$, then $\chi_{\hat\sigma}$ is of $(2w_A-1)\bm J$-positive type.
By Eq.~\eqref{eq:NW_positive_type}, this is precisely the statement that $2w_A-1\in\N(\hat\sigma)$, which proves \cref{thm:single_mode}. 
The symmetry discussed above becomes the symmetry of the NW spectrum: $1-2w_A=2w_B-1$ belongs to $\N(\hat\sigma)$ as well, so that it does not matter which party is called $A$.

Two particular values of $w_A$ illustrate the content of the theorem. For a balanced beam-splitter, $w_A=1/2$ and the theorem asserts that $0\in\N(\hat\sigma)$, i.e.\ by Bochner's theorem that the Wigner function of the output mode is non-negative: a PPT input cannot produce a Wigner-negative output through a balanced beam-splitter or, equivalently, Wigner negativity at the output of a balanced beam-splitter certifies NPT entanglement of the input~\cite{zaw2024certifiable}. For $w_A\neq1/2$, the theorem asserts that $W_{\hat\sigma}$ is a bona fide Wigner function for the value $|2w_A-1|$ of Planck's constant, which is smaller than one for passive transformations, where $w_A\in[0,1]$, and can exceed one for active ones, such as the two-mode squeezer of gain $g$, for which $|2w_A-1|=2g-1$.

\subsection{The Gaussian case: proof of \cref{thm:gaussian}}
\label{sec:gaussian_case}

The ``if'' part of \cref{thm:gaussian} is \cref{thm:single_mode}, so we only have to prove the converse: if a Gaussian state is NPT across $A|B$, then some output mode $\hat b$ of some Gaussian unitary has $2w_A-1\notin\N(\hat\sigma)$. The proof has two steps. We first show that, for a Gaussian input, the single-mode test of \cref{thm:single_mode} is nothing but the uncertainty relation of the mirror state $\hat\rho^{\Gamma_B}$ restricted to the mode $\hat b$. We then show that, whenever the mirror state violates the uncertainty relation, one can find a mode on which it does.

Let $\hat\rho$ be Gaussian with covariance matrix $\bm V$; its mean vector plays no role and we set it to zero. By Simon's criterion, Eq.~\eqref{eq:simon}, $\hat\rho$ is PPT across $A|B$ if and only if
\begin{align}
    \bm H\coloneqq\bm V+\tfrac i2\,\bm\Lambda_B\bm\Omega\bm\Lambda_B\succeq0 ,
    \label{eq:H_def}
\end{align}
which is the uncertainty relation of the mirror state, whose commutation relations are reversed on $B$ (\cref{sec:background_PPT}).

Consider now an output mode $\hat b$ of a Gaussian unitary with symplectic matrix $\bm S$, and let $\bm S_b$ be the $2\times2N$ block of $\bm S$ formed by the rows of $\hat x_b$ and $\hat p_b$, so that $(\hat x_b,\hat p_b)^T=\bm S_b\hat{\mathbf r}$. The reduced state $\hat\sigma$ of $\hat b$ is Gaussian with covariance matrix $\bm V_b=\bm S_b\bm V\bm S_b^T$. Moreover, we have that $\bm S_b\bm\Omega\bm S_b^T=\bm J$, which is the commutator $[\hat b,\hat b^\dagger]=1$, and, by Eqs.~\eqref{eq:reversed_share} and \eqref{eq:weights}, $\bm S_b\bm\Lambda_B\bm\Omega\bm\Lambda_B\bm S_b^T=(w_A-w_B)\bm J=(2w_A-1)\bm J$, which is the commutator that $\hat b$ would have in the mirror state. Therefore,
\begin{align}
    \bm S_b\,\bm H\,\bm S_b^T=\bm V_b+\tfrac i2\,(2w_A-1)\,\bm J .
    \label{eq:gaussian_compression}
\end{align}
The right-hand side is the uncertainty relation of $\hat\sigma$ with $\hbar$ replaced by $2w_A-1$. For a Gaussian state, this is precisely the condition for $2w_A-1$ to belong to the NW spectrum: as recalled in \cref{sec:TMSentdet}, $\eta\in\N(\hat\sigma)$ if and only if $\bm V_b+\frac{i\eta}2\bm J\succeq0$, i.e.\ $\N(\hat\sigma)=[-2\nu_b,2\nu_b]$ with $\nu_b=\sqrt{\det\bm V_b}$. We thus arrive at
\begin{align}
    2w_A-1\in\N(\hat\sigma)
    \quad\Longleftrightarrow\quad
    \bm S_b\,\bm H\,\bm S_b^T\succeq0 .
    \label{eq:gaussian_test}
\end{align}
In words: for a Gaussian input, testing whether $2w_A-1$ belongs to $\N(\hat\sigma)$ amounts to testing whether the mirror state satisfies the uncertainty relation on the mode $\hat b$ alone. The ``if'' part of \cref{thm:gaussian} is also visible here: if $\bm H\succeq0$, every compression $\bm S_b\bm H\bm S_b^T$ is positive semidefinite, so that no mode can fail the test.

Suppose now that $\hat\rho$ is NPT, so that $\bm H$ is not positive semidefinite: there exists $z=a+ib\in\complex^{2N}$, with $a,b\in\reals^{2N}$, such that $z^\dagger\bm Hz<0$. We want to turn $z$ into a mode. Let $\bm Q$ be the real $2\times2N$ matrix with rows $a^T$ and $b^T$. Then $z=\bm Q^Tc$ with $c=(1,i)^T$, so that $c^\dagger(\bm Q\bm H\bm Q^T)c=z^\dagger\bm Hz<0$: the $2\times2$ matrix $\bm Q\bm H\bm Q^T$ is not positive semidefinite. If the rows of $\bm Q$ were the quadratures of a mode, that is, if $\bm Q\bm\Omega\bm Q^T=\bm J$, we would be done: $\bm Q$ would be the block $\bm S_b$ of some symplectic matrix, and Eq.~\eqref{eq:gaussian_test} would give $2w_A-1\notin\N(\hat\sigma)$. In general, however, $\bm Q\bm\Omega\bm Q^T=\omega\bm J$ with $\omega=a^T\bm\Omega b$, the commutator of the observables $a^T\hat{\mathbf r}$ and $b^T\hat{\mathbf r}$, which need not equal one and may even vanish. Two adjustments take care of this.

First, if $\omega=0$, we replace $b$ by $b+\epsilon\,\bm\Omega a$. Notice that $a\neq0$, for otherwise $z^\dagger\bm Hz=b^T\bm Vb\ge0$; hence $a^T\bm\Omega(\bm\Omega a)=-|a|^2\neq0$, so that the new commutator is nonzero, while the strict inequality $z^\dagger\bm Hz<0$ survives for $\epsilon$ small enough. Second, once $\omega\neq0$, we exchange $a$ and $b$ if $\omega<0$, which changes its sign and only permutes the rows and columns of $\bm Q\bm H\bm Q^T$, and we set $\bm S_b\coloneqq\bm Q/\sqrt\omega$. Then $\bm S_b\bm\Omega\bm S_b^T=\bm J$: the rows of $\bm S_b$ are the quadratures of a mode $\hat b$, and they can be completed into a symplectic matrix $\bm S$~\cite{arvind1995real}, so that $\hat b$ is an output mode of a Gaussian unitary. Finally, $\bm S_b\bm H\bm S_b^T=\bm Q\bm H\bm Q^T/\omega$ is not positive semidefinite, and Eq.~\eqref{eq:gaussian_test} gives $2w_A-1\notin\N(\hat\sigma)$. This proves \cref{thm:gaussian}.

Let us remark that the first adjustment cannot be dispensed with. For a two-mode squeezed vacuum, the natural choice for $z$ is given by the EPR combinations $\hat x_1-\hat x_2$ and $\hat p_1+\hat p_2$: their variances $e^{-2r}$ violate the uncertainty relation of the mirror state, but they commute, so they do not form a mode. Any nearby pair with a nonzero commutator does, which is why an arbitrarily weak or arbitrarily unbalanced two-mode squeezer suffices to detect the entanglement of a two-mode squeezed vacuum (\cref{sec:TMSentdet}).

\section{Gabor-frame form of the KLM conditions}
\label{app:gabor}

In this section, we provide the derivations for the results in \cref{sec:testNW} of the main text.

\medskip

We follow Ref.~\cite{cordero2019}, writing the symplectic form as $-z^T \bm{\Omega} z' = p x' - x p'$ for $z = (x,p)$ and $z' = (x',p')$. For $\eta > 0$ and $\psi, \varphi \in L^2(\reals)$, let
\begin{align}
    W_\eta(\psi,\varphi)(x,p) = \frac{1}{2\pi\eta} \int \mathrm{d}y \ e^{-\frac{i}{\eta} p y} \, \psi\!\left(x + \tfrac{y}{2}\right) \overline{\varphi\!\left(x - \tfrac{y}{2}\right)} = \frac{1}{\eta} \, W_1(\psi,\varphi)\!\left(x, \frac{p}{\eta}\right),
    \label{eq:app_Weta}
\end{align}
where $W_1(\psi,\varphi)$ is the ordinary Wigner function of the rank-1 operator $\ket{\psi}\!\bra{\varphi}$. Given a real phase-space function $F$, the operator $\hat{A}_\eta[F]$ is defined weakly by
\begin{align}
    \bra{\varphi} \hat{A}_\eta[F] \ket{\psi} = 2\pi\eta \int \mathrm{d}^2 z \ F(z) \, W_\eta(\psi,\varphi)(z).
    \label{eq:app_Aeta}
\end{align}
By the second equality in Eq.~\eqref{eq:app_Weta} and the overlap formula $\Tr[\hat{X} \hat{Y}] = 2\pi \int W_{\hat{X}} W_{\hat{Y}}$, the Wigner function of $\hat{A}_\eta[F]$ is $\eta \, F(x, \eta p)$, i.e.\ $F$ compressed by a factor $\eta$ along $p$; in particular $\hat{A}_1[F]$ is the operator with Wigner function $F$, $\Tr \hat{A}_\eta[F] = \int F$, and $\Tr \hat{A}_\eta[F]^2 = 2\pi\eta \int F^2$. The NW spectrum of $\hat{\sigma}$ is the set of $\eta$ such that $\hat{A}_{|\eta|}[W_{\hat{\sigma}}] \succeq 0$ \cite{narcowich_conditions_1988} (see \cref{app:NWproperties}); we write $\hat{A}_\eta[\hat{\sigma}]$ for $\hat{A}_\eta[W_{\hat{\sigma}}]$. Since $W_{\hat{\sigma}}$ is square integrable, $\hat{A}_\eta[\hat{\sigma}]$ is Hilbert--Schmidt, hence bounded. Ref.~\cite{cordero2019} omits the factor $2\pi\eta$ in Eq.~\eqref{eq:app_Aeta}, which does not affect positivity.

\emph{Displaced windows and Gabor frames.} For $\lambda = (\lambda_x, \lambda_p) \in \reals^2$, let $\hat{T}_\eta(\lambda) = \hat{D}(\lambda_x, \lambda_p/\eta)$, i.e.
\begin{align}
    \big( \hat{T}_\eta(\lambda) \psi \big)(x) = e^{\frac{i}{\eta} \left( \lambda_p x - \frac{1}{2} \lambda_p \lambda_x \right)} \, \psi(x - \lambda_x),
    \label{eq:app_Teta}
\end{align}
the displacement operator to the phase-space point $(\lambda_x, \lambda_p/\eta)$; it satisfies $\hat{T}_\eta(\lambda) \hat{T}_\eta(\mu) = e^{-\frac{i}{2\eta} \lambda^T \bm{\Omega} \mu} \hat{T}_\eta(\lambda + \mu)$. Given a normalized window $\phi \in L^2(\reals)$ and a lattice $\Lambda \subset \reals^2$, let $\phi_\lambda = \hat{T}_\eta(\lambda) \phi$. The family $\{\phi_\lambda\}_{\lambda \in \Lambda}$ is a Gabor frame if there exist $0 < A \leq B < \infty$ such that
\begin{align}
    A \| \psi \|^2 \leq \sum_{\lambda \in \Lambda} |\braket{\phi_\lambda | \psi}|^2 \leq B \| \psi \|^2 \qquad \forall \psi \in L^2(\reals).
    \label{eq:app_frame}
\end{align}
The lower bound guarantees that no nonzero vector is orthogonal to all the $\phi_\lambda$, so that their finite linear combinations are dense in $L^2(\reals)$; in general the family is overcomplete, and the expansion coefficients of a given vector are not unique. For a Gaussian window $\phi(x) \propto e^{-x^2/2w^2}$ and $\Lambda = a\mathbb{Z} \times b\mathbb{Z}$, the frame property holds if and only if $ab < 2\pi\eta$, for any width $w$ \cite{lyubarskii1992,seip1992}: the lattice spacings and the width of the window are independent parameters.

\emph{The matrix $\bm{K}_F$.} For any finite $F \subset \Lambda$ and coefficients $c \in \complex^F$, the vector $\psi_c = \sum_{\lambda \in F} c_\lambda \phi_\lambda$ satisfies
\begin{align}
    \bra{\psi_c} \hat{A}_\eta[\hat{\sigma}] \ket{\psi_c} = \sum_{\lambda, \mu \in F} \bar{c}_\lambda \, \bra{\phi_\lambda} \hat{A}_\eta[\hat{\sigma}] \ket{\phi_\mu} \, c_\mu = c^\dagger \bm{K}_F \, c,
\end{align}
so that $\bm{K}_F \succeq 0$ if and only if $\hat{A}_\eta[\hat{\sigma}]$ is nonnegative on the span of $\{\phi_\lambda\}_{\lambda \in F}$. If $\hat{A}_\eta[\hat{\sigma}] \succeq 0$, all the $\bm{K}_F$ are therefore positive semidefinite. Conversely, if $\bm{K}_F \succeq 0$ for every finite $F$, then $\hat{A}_\eta[\hat{\sigma}]$ is nonnegative on the finite linear combinations of the frame vectors, which are dense, hence on the whole Hilbert space by continuity of the quadratic form of a bounded operator. This proves the equivalence stated in \cref{sec:testNW}, which is Theorem 6 of Ref.~\cite{cordero2019} (there, the matrix is $\bm{M}_F = \bm{K}_F^T$). Note that the frame property is only needed for the converse, and that $\bm{K}_F$ is not the matrix of $\hat{A}_\eta[\hat{\sigma}]$ in an orthonormal basis but rather a Gram-type matrix, $\bm{K}_F = \bm{T}_F^\dagger \hat{A}_\eta[\hat{\sigma}] \bm{T}_F$ with $\bm{T}_F : c \mapsto \psi_c$ the synthesis map of the finite frame.

\emph{Entries.} Direct substitution of Eq.~\eqref{eq:app_Teta} into Eq.~\eqref{eq:app_Weta} gives $W_\eta(\phi_\lambda, \phi_\mu)(z) = e^{\frac{i}{2\eta} \lambda^T \bm{\Omega} \mu} \, e^{\frac{i}{\eta} z^T \bm{\Omega} (\lambda - \mu)} \, W_\eta \phi\big( z - \frac{\lambda + \mu}{2} \big)$, with $W_\eta \phi \coloneqq W_\eta(\phi,\phi)$: the midpoint $(\lambda + \mu)/2$ shifts the envelope and the separation $\lambda - \mu$ sets the oscillating phase. Hence, by Eq.~\eqref{eq:app_Aeta},
\begin{align}
    \bra{\phi_\lambda} \hat{A}_\eta[\hat{\sigma}] \ket{\phi_\mu} = 2\pi\eta \, e^{\frac{i}{2\eta} \mu^T \bm{\Omega} \lambda} \int \mathrm{d}^2 z \ W_{\hat{\sigma}}(z) \, e^{\frac{i}{\eta} z^T \bm{\Omega} (\mu - \lambda)} \, W_\eta \phi\!\left( z - \frac{\lambda + \mu}{2} \right).
    \label{eq:app_entries}
\end{align}
For the Gaussian window of width $w$, $W_\eta \phi(z) = \frac{1}{\pi\eta} \, e^{-x^2/w^2 - w^2 p^2/\eta^2}$.

\emph{Measurement of the entries.} By the second equality in Eq.~\eqref{eq:app_Weta}, Eq.~\eqref{eq:app_Aeta} can be rewritten as
\begin{align}
    \bra{\phi_\lambda} \hat{A}_\eta[\hat{\sigma}] \ket{\phi_\mu} = 2\pi \int \mathrm{d}x \, \mathrm{d}p \ W_{\hat{\sigma}}(x,p) \, W_1\big(\phi_\mu, \phi_\lambda\big)\!\left(x, \frac{p}{\eta}\right) = \Tr\big[ \hat{\sigma} \, \hat{O}_{\lambda\mu} \big],
    \label{eq:app_Olm}
\end{align}
where $\hat{O}_{\lambda\mu}$ is the operator whose Wigner function is that of $\ket{\phi_\mu}\!\bra{\phi_\lambda}$ stretched by a factor $\eta$ along $p$. The $\eta$-dependence is thus entirely moved onto the observables $\hat{O}_{\lambda\mu}$, and the question is which measurements give access to their expectation values. For the Gaussian window of width $w$, the stretching maps the centre $(\lambda_x, \lambda_p/\eta)$ of $\phi_\lambda$ back to $\lambda$, so that $\hat{O}_{\lambda\lambda}/\eta$ has a Gaussian Wigner function centred at $\lambda$ with variances $w^2/2$ along $x$ and $\eta^2/2w^2$ along $p$. Their product is $\eta^2/4$ for any $w$: for $\eta < 1$, $\hat{O}_{\lambda\lambda}$ violates the uncertainty relation and is not proportional to a state, whereas for $\eta > 1$ it is proportional to a mixed Gaussian state, which is the displaced thermal state with mean photon number $(\eta-1)/2$ for $w^2 = \eta$. Heterodyne detection samples the Husimi function $Q_{\hat{\sigma}}(z) = \bra{z} \hat{\sigma} \ket{z}/2\pi$, with $\ket{z}$ the coherent state centred at $z$, so that for any bounded function $g$ of the outcomes
\begin{align}
    \int \mathrm{d}^2 z \ Q_{\hat{\sigma}}(z) \, g(z) = \Tr\big[ \hat{\sigma} \hat{G} \big], \qquad \hat{G} = \int \frac{\mathrm{d}^2 z}{2\pi} \ g(z) \ket{z}\!\bra{z},
    \label{eq:app_hetmean}
\end{align}
i.e.\ the sample mean of $g$ over the heterodyne outcomes is an unbiased estimator of $\Tr[\hat{\sigma} \hat{G}]$. The expectation value of $\hat{O}_{\lambda\mu}$ is thus obtained in this way if and only if $\hat{O}_{\lambda\mu} = \hat{G}$ for some bounded $g$, that is, since the Wigner function of $\ket{z}\!\bra{z}$ is the vacuum Gaussian $G_{\mathrm{vac}}(z') = \frac{1}{\pi} e^{-|z' - z|^2}$ centred at $z$, if and only if
\begin{align}
    W_{\hat{O}_{\lambda\mu}} = g * G_{\mathrm{vac}}
    \label{eq:app_deconv}
\end{align}
has a bounded solution $g$. For a Gaussian $\hat{O}_{\lambda\mu}$ this is the case if and only if both variances of $\hat{O}_{\lambda\mu}$ exceed the vacuum value $1/2$, i.e.\ $1 < w^2 < \eta^2$: for $\eta > 1$ this is achieved by any window squeezed along $p$ with $1 < w < \eta$ (the vacuum window is excluded, its variance along $x$ being exactly $1/2$), while for $\eta \le 1$ no Gaussian window works, nor does unbalanced heterodyne detection, whose projectors are squeezed coherent states with the same product of variances. The choice $w^2 = \eta$ minimises the supremum of $g$; for it, solving Eq.~\eqref{eq:app_deconv} with $W_{\hat{O}_{\lambda\mu}}$ read from Eq.~\eqref{eq:app_entries} gives Eq.~\eqref{eq:gabor_heterodyne} with
\begin{align}
    g_{\lambda\mu}(z) = \frac{2\eta}{\eta - 1} \, e^{\frac{i}{2\eta} \mu^T \bm{\Omega} \lambda} \exp\!\left[ i \kappa \cdot c + \frac{i\eta}{\eta - 1} \, \kappa \cdot (z - c) + \frac{\eta}{4(\eta - 1)} |\kappa|^2 - \frac{|z - c|^2}{\eta - 1} \right],
    \qquad c = \frac{\lambda + \mu}{2}, \quad \kappa = \frac{1}{\eta} \bm{\Omega} (\mu - \lambda),
    \label{eq:app_g}
\end{align}
where $\kappa \cdot z = z^T \bm{\Omega} (\mu - \lambda)/\eta$ is the phase appearing in Eq.~\eqref{eq:app_entries}; for $\lambda = \mu$, $g_{\lambda\lambda}(z) = \frac{2\eta}{\eta-1} e^{-|z-\lambda|^2/(\eta-1)}$ is $\eta$ times the weight with which the displaced thermal state $\hat{O}_{\lambda\lambda}/\eta$ is written as a mixture of coherent states. The lattice enters only through the functions $g_{\lambda\mu}$ that are averaged, never through a discretisation of the outcomes. Since $|g_{\lambda\mu}(z)| \leq \frac{2\eta}{\eta - 1} \, e^{|\lambda - \mu|^2 / 4\eta(\eta - 1)}$, by Hoeffding's inequality each entry is estimated within $\epsilon$ with probability $1 - \delta$ from $O\big(\epsilon^{-2} \log(1/\delta)\big)$ heterodyne samples, the same samples serving for all entries and all $F$; the cost grows with the distance $|\lambda - \mu|$ and as $\eta \to 1^+$. For $\eta < 1$ and a Gaussian window the characteristic function of $\hat{O}_{\lambda\mu}$ is still a Gaussian, so that the pattern functions of phase-randomised homodyne detection are bounded \cite{lvovsky2009continuous} and the entries can be estimated without bias; from heterodyne data they can only be estimated with a regularisation bias, along the lines of Ref.~\cite{chabaud2020building,chabaud2020certification}.

\section{Proof of \cref{thm:negativity}}
\label{app:quantitative}

In this section we give a short proof of \cref{thm:negativity} in the main text.

\medskip

By Eq.~\eqref{eq:bridge}, $\chi_{\hat{\sigma}}(x) = \Tr[\hat{\rho}^{\Gamma_B} e^{i \hat{q}'^T_O \bm{J} x}]$, where the quadratures $\hat{q}'_O$ of the output mode built from the reflected input quadratures satisfy $[\hat{q}'_O, \hat{q}'^T_O] = i \eta \bm{\Omega}$ by Eqs.~\eqref{eq:reversed_share} and \eqref{eq:weights}. 

The operators $\hat{V}_j = e^{i \hat{q}'^T_O \bm{J} x_j}$ are unitary and obey the Weyl relation \eqref{eq:weyl} with $\bm{K} = \eta \bm{J}$, so that
\begin{align}
    \bm{M}_{jk} = \Tr\big[ \hat{\rho}^{\Gamma_B} \, \hat{V}_j \hat{V}_k^\dagger \big].
\end{align}
For $c \in \complex^m$ let $\hat{Y}_c = \sum_j \bar{c}_j \hat{V}_j$, so that $\Tr[\hat{\rho}^{\Gamma_B} \hat{Y}_c \hat{Y}_c^\dagger] = c^\dagger \bm{M} c$ and, by unitarity, $\| \hat{Y}_c \|_\infty \leq \| c \|_1 \leq \sqrt{m} \, \| c \|_2$. Let $v^{(1)}, \dots, v^{(m)}$ be an orthonormal eigenbasis of $\bm{M}$ with eigenvalues $\lambda_1, \dots, \lambda_m$. Since $\sum_\ell v^{(\ell)} v^{(\ell)\dagger} = \bm{1}$,
\begin{align}
    \sum_{\ell = 1}^m \hat{Y}_{v^{(\ell)}} \hat{Y}_{v^{(\ell)}}^\dagger = \sum_{j,k} \Big( \sum_\ell \bar{v}^{(\ell)}_j v^{(\ell)}_k \Big) \hat{V}_j \hat{V}_k^\dagger = \sum_j \hat{V}_j \hat{V}_j^\dagger = m \, \hat{1},
\end{align}
hence the operator $\hat{Q}_- = \frac{1}{m} \sum_{\ell : \lambda_\ell < 0} \hat{Y}_{v^{(\ell)}} \hat{Y}_{v^{(\ell)}}^\dagger$ satisfies $0 \preceq \hat{Q}_- \preceq \hat{1}$ and $\Tr[\hat{\rho}^{\Gamma_B} \hat{Q}_-] = \frac{1}{m} \sum_{\lambda_\ell < 0} \lambda_\ell = - \Tr \bm{M}_- / m$. Write $\hat{\rho}^{\Gamma_B} = \hat{X}_+ - \hat{X}_-$ with $\hat{X}_\pm \succeq 0$ orthogonal, so that $\Tr \hat{X}_- = N(\hat{\rho})$ (if $\hat{\rho}^{\Gamma_B}$ is not trace-class, $N(\hat{\rho}) = \infty$ and there is nothing to prove). Then
\begin{align}
    \frac{\Tr \bm{M}_-}{m} = \Tr[\hat{X}_- \hat{Q}_-] - \Tr[\hat{X}_+ \hat{Q}_-] \leq \Tr[\hat{X}_- \hat{Q}_-] \leq \Tr \hat{X}_- = N(\hat{\rho}),
\end{align}
which is the first bound in Eq.~\eqref{eq:negativity_bound}; the second follows from $\Tr \bm{M}_- \geq \max\{0, -\lambda_{\min}(\bm{M})\}$, and the third from $\| \bm{M} \|_1 = m + 2 \Tr \bm{M}_-$ and $\| \hat{\rho}^{\Gamma_B} \|_1 = 1 + 2 N(\hat{\rho})$. If a single normalised eigenvector $v$ with $v^\dagger \bm{M} v = \lambda_{\min}(\bm{M}) < 0$ is used, the operator $\hat{Y}_v \hat{Y}_v^\dagger / \| v \|_1^2$ lies between $0$ and $\hat{1}$ and the same argument gives $N(\hat{\rho}) \geq -\lambda_{\min}(\bm{M}) / \| v \|_1^2$, which is sharper since $\| v \|_1^2 \leq m$. Notice that the Weyl relation is all that was used, so that the proof holds for $\eta = 0$ as well, and that nothing was assumed on $\hat{\rho}$.

For the matrices $\bm{K}_F$ of \cref{sec:testNW}, the unitaries $\hat{V}_j$ are replaced by vectors. Since the $\hat{V}_j$ obey the Weyl relation with parameter $\eta$, by the Stone--von Neumann theorem the Hilbert space of the input modes decomposes as $L^2(\reals) \otimes \mathcal{K}$ with $e^{i x^T \hat{q}'_O} = \hat{T}_\eta(\lambda_x) \otimes \hat{1}$ for a suitable linear relabeling $x \mapsto \lambda_x$ of the points; the operator $\Tr_{\mathcal{K}}[\hat{\rho}^{\Gamma_B}]$ on $L^2(\reals)$ then has the same matrix elements $\bra{\phi_\lambda} \cdot \ket{\phi_\mu}$ as $\hat{A}_\eta[\hat{\sigma}]$, by Eq.~\eqref{eq:app_Aeta}, and $\| \Tr_{\mathcal{K}}[\hat{\rho}^{\Gamma_B}] \|_1 \leq \| \hat{\rho}^{\Gamma_B} \|_1$. The argument above with $\hat{Y}_c = \sum_\lambda c_\lambda \ket{\phi_\lambda}$ and $\sum_\ell \hat{Y}_{v^{(\ell)}} \hat{Y}_{v^{(\ell)}}^\dagger = \sum_\lambda \ket{\phi_\lambda}\!\bra{\phi_\lambda} \preceq \| \bm{G}_F \| \, \hat{1}$, with $\bm{G}_F$ the Gram matrix of the $\phi_\lambda$, gives $N(\hat{\rho}) \geq \Tr (\bm{K}_F)_- / \| \bm{G}_F \|$.

For the example of \cref{sec:quantitative}, $\chi_{\hat{\sigma}}(x) = e^{-|x|^2/4}$ and, with $a = e^{-s^2/4}$ and $\theta = \eta s^2 / 2$, the three-point matrix and its determinant are
\begin{align}
    \bm{M} =
    \begin{pmatrix}
        1 & a & a \\
        a & 1 & a^2 e^{i\theta} \\
        a & a^2 e^{-i\theta} & 1
    \end{pmatrix},
    \qquad
    \det \bm{M} = (1 - a^2)^2 - 2 a^4 (1 - \cos\theta) = \frac{1 - \eta^2}{4} \, s^4 + O(s^6),
\end{align}
which is negative for small $s$ whenever $\eta > 1$, i.e.\ for any $r > 0$. For $r = 1/2$ and $s = 1$, $\lambda_{\min}(\bm{M}) \approx -0.0351$.

\end{document}